\documentclass[11pt,a4paper]{article}
\usepackage{geometry}
\usepackage[normalem]{ulem}
\usepackage{microtype}           % Better typography
\usepackage{amsmath,amssymb,amsthm}
\usepackage{mathtools}
\mathtoolsset{showonlyrefs}
\usepackage{bm}
\usepackage{graphicx}
\usepackage{subfigure}
\usepackage{caption}
\usepackage{mathrsfs}
\usepackage{amsthm}
\usepackage[numbers,square,sort&compress]{natbib}
\usepackage[colorlinks=true,linkcolor=black,citecolor=blue,urlcolor=black,bookmarks]{hyperref}
\usepackage{xcolor}
\usepackage{todonotes}

\newcommand{\affiliation}[1]{\\ \parbox[t]{0.9\textwidth}{\centering\small\textit{#1}}}  %Affiliation with auto wrap and italics
\def \be {\begin{equation}}
\def \ee {\end{equation}}
\def \ba {\begin{array}}
\def \ea {\end{array}}
\def \bea{\begin{eqnarray}}
\def \eea{\end{eqnarray}}
\def \nn {\nonumber}

\numberwithin{equation}{section}
\newcommand{\ii}{\mathrm i}

\newcommand{\Tr}{\operatorname{Tr}}

\newcommand{\id}{\mathbf1}
\newcommand{\ket}[1]{\lvert#1\rangle}
\newcommand{\bra}[1]{\langle#1\rvert}
\newcommand{\braket}[2]{\langle#1|#2\rangle}

\newcommand{\cH}{\mathcal H}

\theoremstyle{definition}

\title{\vspace{-2cm}\begin{flushright}
{\footnotesize{YITP-26-165}}
\end{flushright}
\vspace{2cm}
Two Faces of Spacetime Density Matrices:\\
Entanglement and Complexity}
\author{Wu-zhong Guo\footnote{wuzhong@hust.edu.cn}~~\textsuperscript{$\otimes$,$\circledast$}
\affiliation{\textsuperscript{$\otimes$}School of Physics, Huazhong University of Science and Technology, \\
Luoyu Road 1037, Wuhan, Hubei 430074, China
}
\affiliation{\textsuperscript{$\circledast$}Center for Gravitational Physics and Quantum Information,\\
Yukawa Institute for Theoretical Physics, 
Kyoto University, Kyoto 606-8502, Japan}
}
\date{}

\begin{document}

\maketitle

\begin{abstract}
We establish a direct connection between entanglement and complexity through spacetime density matrices. For stationary pure states and complementary regions at different times, spacetime replica moments encode the return amplitude of a replica-defect quench. This representation defines an associated state spread complexity, which we call \textit{Krylov–Rényi complexity}, is equivalently obtained by using a replica-defect excitation of the replicated background state as the seed and evolving it under the original replicated Hamiltonian. We further construct the \textit{Krylov–von Neumann complexity} as a probe limit from first-order replica data. Timelike entanglement and complexity provide complementary information at a fixed time, while the full time dependence of the timelike entropy determines the Krylov dynamics and hence the complexity. The replica construction fixes the seed and its evolution through the background state, subsystem geometry, and regulator, without an independently chosen probe operator. Like entanglement entropy, the resulting complexity is defined nonperturbatively and explore universal property and dynamical information about the underlying theory.
\end{abstract}
\clearpage
\tableofcontents

\section{Introduction}
Entanglement and complexity provide different perspectives on quantum systems. Entanglement plays a central role in quantum information, the characterization of quantum phases, and holographic descriptions of spacetime and black holes, see the reviews \cite{RevModPhys.81.865,RevModPhys.80.517,Takayanagi:2025ula}. Complexity has been used to investigate quantum chaos, information spreading, quantum system dynamics, see \cite{Nandy:2024evd,Chapman:2021jbh,Jeong:2026gdc} for the reviews. Although these two concepts are expected to be related in specific settings, identifying a common physical framework that directly connects entanglement observables with dynamical complexity remains an important goal.

For a bipartite pure state, entanglement is characterized by the von Neumann and Rényi entropies of a reduced density matrix. Replica methods in quantum field theory and geometric prescriptions in holography provide systematic ways to compute entanglement entropy~\cite{Calabrese:2004eu,Ryu:2006bv}. While Complexity admits several inequivalent definitions. Distinct approaches to complexity have been proposed
in quantum field theory and holography~\cite{Stanford:2014jda,
Brown:2015bva,Jefferson:2017sdb,Chapman:2017rqy,Caputa:2017urj,Belin:2021bga,Parker:2018yvk,Balasubramanian:2022tpr}.  Among them, Krylov complexity measures spreading under time evolution in an orthonormal basis generated by repeated action of the Hamiltonian on a seed state, or of the Liouvillian on a seed operator~\cite{Parker:2018yvk,Balasubramanian:2022tpr}. Conventional entanglement refers to a spatial partition at a common time, whereas Krylov complexity explicitly depends on a seed and its evolution\footnote{Entanglement entropy also probes quantum dynamics, for example through its evolution following a quantum quench~\cite{Calabrese:2005in}. In such studies, the entropy is evaluated for a spatial bipartition at each time.}. From this perspective, entanglement and complexity appear to probe different aspects of a quantum system, and a direct connection between them is not immediately apparent. Possible connections have nevertheless been explored through Krylov complexity generated by modular Hamiltonian evolution~\cite{Caputa:2023vyr}.

To investigate this connection under physical time evolution,
we turn to the spacetime density matrix (STDM) framework. Various approaches have been developed to describe temporal correlations and quantum states across different times, extending the conventional density matrix
framework~\cite{Leifer:2013ghr,Fitzsimons:2013gga,
Cotler:2017anu,Diaz:2020dfe,Fullwood:2022rjd}.
More recently, entanglement between time-separated
subsystems can be well-defined and explored using generalized
density operators~\cite{Doi:2022iyj,Milekhin:2025ycm,
Guo:2025dtq,Das:2025fcd,Diaz:2026qrf,Jia:2026iqa}.
In this work, we follow the spacetime density matrix formulation of~\cite{Milekhin:2025ycm,Guo:2025dtq}, which encodes
correlations between subsystems at different times with
a prescribed operator ordering. Timelike entanglement defined in this framework encodes
dynamical information about the system, motivating us
to investigate its connection with Krylov complexity.

 Let$C_0$ and$C_t$ denote the complete system at times$0$ and$t$, respectively. Given an initial density matrix$\rho$ and unitary evolution$U(t)=e^{-iHt}$, the full spacetime density matrix$T_{C_0C_t}$, acting on$\mathcal H_{0}\otimes\mathcal H_{t}$, is defined by
\bea
 \Tr_{\cH_0\otimes\cH_t}
 \!\left[T_{C_0C_t}(X\otimes Y)\right]
 =\Tr_{\cH}\!\left[\rho XU^\dagger(t)YU(t)\right]
 \label{eq:stdm-full-definition}
\eea
for arbitrary operators$X$ and$Y$ on the complete system. The ordering on the right-hand side is part of the definition. To retain only subsystems$A_0\subset C_0$ and$B_t\subset C_t$, we trace out their respective complements, obtaining the reduced spacetime density matrix
\bea
T_{A_0B_t}
=
\operatorname{Tr}_{\bar A_0\,\bar B_t}T_{C_0C_t},
\label{eq:stdm-reduced-definition}
\eea
where $\bar A_0$ and $\bar B_t$ are complementary of the subsystems.
This operator reproduces the ordered correlations between observables supported on the subsystems $A_0$ and $B_t$. For $n$ being integers, the timelike Rényi entropy is defined as\footnote{We use ``timelike entanglement entropy'' broadly:
the regions $A_0$ and $B_t$ may be either spacelike or
timelike separated. For spacelike-separated regions in a
relativistic theory, microcausality ensures that their
observables commute, and $T_{A_0B_t}$ reduces to the
ordinary positive semidefinite reduced density operator
for their union on a common Cauchy surface. The definition
then gives the usual R\'enyi entropy. For timelike-separated
regions, $T_{A_0B_t}$ is generally non-Hermitian, and the
corresponding entropy can be complex.}
\bea
\begin{aligned}
S_n^{\mathrm{time}}(A,B;t)=
\frac{1}{1-n}
\operatorname{Log}\operatorname{Tr}\!\left(T_{A_0B_t}^{\,n}\right).
\end{aligned}
\eea
%The logarithm is specified by continuation along a path on which
%$M_n\ne0$. No purity or stationarity assumption
%was needed for the definitions above.
The timelike entropy can be defined as von Neumann entropy  $$S^{\mathrm{time}}(A,B;t)=
-\operatorname{Tr}\![
T_{A_0B_t}\operatorname{Log}T_{A_0B_t}].$$
%Although$T_{A_0B_t}$ has unit trace, it is generally non-Hermitian, so these quantities can be complex and require specified logarithm branches. The definitions are understood at a finite regulator, with suitable ultraviolet regularization in quantum field theory. 
The entropy defined in this way is also expected to be consistent with the holographic description within the AdS/CFT correspondence~\cite{Doi:2022iyj,Heller:2024whi}.

Timelike entropies encode dynamical information, motivating us to connect them with the spreading of quantum states. Our starting point is Cardy's replica-quench construction, which introduces a defect Hamiltonian by reconnecting couplings between replicas across a spatial partition~\cite{Cardy:2011zz}. For stationary pure states and complementary regions at different times, we find an exact connection between spacetime replica moments and the return amplitude of a specified defect quench. The resulting Hamiltonian and initial state then provide the ingredients for the standard state spread construction~\cite{Balasubramanian:2022tpr}, see Section~\ref{sec:renyi-krylov}. We call the associated replica-dependent complexity the \textit{Krylov–Rényi complexity}, denoted by $\mathcal C_n(t)$. The corresponding \textit{Krylov–von Neumann complexity} is defined as the limit$\mathcal C_{\mathrm{vN}}(t):=\lim_{n\to 1}\frac{\mathcal C_n(t)}{n-1}$.

This establishes a common dynamical framework in which the meanings of timelike entanglement and Krylov complexity can be constructed and compared directly. This framework clarifies the relation between timelike entanglement and Krylov–Rényi complexity in two respects:
\begin{itemize}
\item At a fixed time, the two quantities probe complementary aspects of the defect-quench dynamics: timelike Rényi entanglement encodes the amplitude to return to the initial state, whereas Krylov–Rényi complexity measures the average depth of the state's spreading away from the initial state along the Krylov chain.
\item The full time dependence of the timelike Rényi quantity, including both its real and imaginary parts, determines the Krylov dynamics and allows the complexity to be reconstructed for a fixed preparation.
\end{itemize}

We illustrate how to evaluate Krylov-R\'enyi and Krylov-von Neumann compelxity from timelike entanglement with three examples in quantum field theories: a half-line in a 2-dimensional CFT, a finite interval in the massless Dirac theory, and a planar cut in a 3-dimensional CFT.
The half-line gives exact quadratic growth of the Krylov--R\'enyi complexity.
The Dirac example reveals the effect of a finite interval:
the timelike entropy approaches a constant, while a
semiclassical analysis predicts quadratic late-time growth
of the Krylov--von Neumann complexity.
For the planar cut, the spatial strip entropy determines
the first-order replica data, yielding a closed expression
for the complexity density with a logarithmic correction
to its leading quadratic growth.

The paper is organized as follows. Section~\ref{sec:sewing} introduces spacetime density matrices and derives the replica-defect representation of their moments. Section~\ref{sec:krylov} defines Krylov–Rényi and Krylov–von Neumann complexity, develops the spectral reconstruction and field-theory regularization, and relates their short-time expansions to derivatives of timelike entropies. Section~\ref{sec:examples} illustrates the framework through a half-line in a 2-dimensional CFT, a finite interval in massless Dirac theory, and a planar cut in a 3-dimensional CFT. Section~\ref{Sec_discussion} summarizes our main results and discusses prospects for numerical calculations of the complexity and its possible holographic dual. The appendices provide further details of the derivations and proofs presented in the main text.

\section{Spacetime density matrices and the replica-defect Hamiltonian}
\label{sec:sewing}
In this section we firstly show the definition of spacetime density matrices and definition of timelike R\'enyi entropy based on it. We then show timelike R\'enyi can be reformulated  as return amplitude with the defect Hamiltonian in the replica $n$ theory. 

\subsection{The spacetime density matrix and timelike R\'enyi}
\label{sec:stdm-definition}
%Let $C_0$ and $C_t$ be complete Cauchy slices at times $0$ and $t$.
%Their Hilbert spaces $\cH_0$ and $\cH_t$ are labeled copies of the
%same regulated Hilbert space $\cH$. Take an initial density operator
%$\rho\ge0$, $\Tr\rho=1$, and let $U(t)=e^{-i Ht}$.
%The spacetime density matrix on the two complete slices is the
%unique operator $T_{C_0C_t}$ on $\cH_0\otimes\cH_t$ satisfying

%for all full-system operators $X$ and $Y$. The ordering on the
%right-hand side is part of the definition. The two tensor factors
%record the two insertion times; they do not represent independently
%prepared systems. Products $X\otimes Y$ span the full operator space,
%so the trace pairing determines $T_{C_0C_t}$ uniquely.

%Choose $C_0=A_0\cup\bar A_0$ and $C_t=B_t\cup\bar B_t$, with
%$\cH_0=\cH_A\otimes\cH_{\bar A}$ and
%$\cH_t=\cH_B\otimes\cH_{\bar B}$.
By definition of STDM (\ref{eq:stdm-full-definition}) and reduced STDM (\ref{eq:stdm-reduced-definition}), for any operators $X_A$ and $X_B$ in subregions $A$ and $B$ respectively, we have
\bea
 \Tr_{\cH_A\otimes\cH_B}
 \!\left[T_{A_0B_t}(X_A\otimes Y_B)\right]
 =\Tr_{\cH}\!\left[
 \rho (X_A\otimes\id_{\bar A})U^\dagger
       (Y_B\otimes\id_{\bar B})U\right].
 \label{eq:stdm}
\eea
 The subsystem Hilbert-space labels refer to their respective time slots.
%Hereafter, an operator with a subsystem label is understood to act as
%the identity on its complement when used inside a full-system trace
%or expectation value. The state $\rho$ and evolution $U$ remain
%full-system objects. Even when $A$ and $B$ overlap spatially, their
%retained factors belong to different time slots.
For later motivation, write
$T_{ab,a'b'}=\langle ab|T_{A_0B_t}|a'b'\rangle$ in orthonormal
subsystem bases $\{ |ab\rangle:=|a\rangle\otimes |b\rangle \} \in \cH_A \otimes \cH_B$ . Inserting matrix units in \eqref{eq:stdm} gives
\begin{equation}
 T_{ab,a'b'}=
 \Tr\!\left[\rho(\ket{a'}\bra a)_A
 U^\dagger(\ket{b'}\bra b)_BU\right],
 \label{eq:stdm-components}
\end{equation}
where, $(\ket{a'}\bra a)_A=\ket{a'}\bra a\otimes\id_{\bar A}$.
%The full environment-index contraction is given in
%Appendix~\ref{app:sewing-details}.
As a check, setting either insertion to the identity yields
\bea
 \Tr T_{A_0B_t}=1,\qquad
 \Tr_B T_{A_0B_t}=\rho_A,\qquad
 \Tr_A T_{A_0B_t}=(U\rho U^\dagger)_B,
 \label{eq:stdm-marginals}
\eea
where $\rho_A=\Tr_{\bar A}\rho$ and
$(U\rho U^\dagger)_B=\Tr_{\bar B}(U\rho U^\dagger)$.

It can be shown the moments of STDM $T_{C_0C_t}$ are time
independent. But the reduced moments can have nontrivial time dependence \cite{Guo:2025dtq}. The STDM formalism thus can provide a framework for understanding entanglement for arbitrary subregions in spacetime.
For later convenience, we define the replica moments for integers $n\ge2$ by
\begin{equation}
 M_n(A,B;t):=\Tr T_{A_0B_t}^{\,n}
 =e^{(1-n)S_n^{\rm time}(A,B;t)}.
 \label{eq:time-renyi}
\end{equation}

\subsection{Moments of reduced STDM and the replica-defect Hamiltonian}

The moments $M_n(A,B;t)$ can be evaluated in the $n$-replicated theory by cyclically sewing the replicas along the retained regions.
In QFTs, this sewing is implemented by replica permutation defects. In two dimensions, the endpoints of the cuts are represented by branch-point twist fields \cite{Cardy:2007mb}.
We next express these moments using regional permutation operators in the replicated theory. 

For a subsystem $R$, define the regional cycle by its action on the replicated subsystem basis:
\begin{equation}
 P_R\ket{r_1,r_2,\ldots,r_n}
 =\ket{r_2,\ldots,r_n,r_1}.
 \label{eq:cycle-action}
\end{equation}
It acts trivially on the complementary factors.
The replica number $n$ is implicit in the notation for all permutation operators.
This convention implies
\begin{equation}
 \Tr_{\cH_R^{\otimes n}}
 \!\left[(X_1\otimes\cdots\otimes X_n)P_R\right]
 =\Tr_R(X_1X_2\cdots X_n).
 \label{eq:cycle-convention}
\end{equation}
All regional cycles below use this orientation. For the replica-defect construction we now specialize to a normalized pure initial state, $\rho=\ket\psi\bra\psi$, without assuming that $\ket\psi$ is an energy eigenstate. 
For the ordered spacetime density matrix \eqref{eq:stdm}, the moments can be formulated as the replica sewing form as follows,
\begin{equation}
 M_n(A,B;t)=
 \bra{\Omega_n}P_A U_n^\dagger P_B U_n\ket{\Omega_n},
 %=\Tr\!\left[\rho_nP_A U_n^\dagger P_B U_n\right].
 \label{eq:replica-identity}
\end{equation}
where $\ket{\Omega_n}:=\ket\psi^{\otimes n}$, and $U_n=U^{\otimes n}=e^{-i H_nt}$ with $H_n=\sum_{a=1}^n H^{(a)}$. Here $a=1,\ldots,n$ labels the replicas. The two regional insertions have the same cycle orientation. No stationarity or locality assumption is required. We prove this formula in Appendix~\ref{app:sewing-details}.

We now specialize to $A=\bar R$ and $B=R$, with
$\cH=\cH_{\bar R}\otimes\cH_R$.
The two regional cycles commute and their product is the full replica cycle:
$ P_{\bar R}P_R=P_{\rm all}$.
Because all replicas contain the same pure state,
$P_{\rm all}\ket{\Omega_n}=\ket{\Omega_n}$.
Consequently,
\bea
 \bra{\Omega_n}P_{\bar R}
 =\bra{\Omega_n}P_{\rm all}P_R^\dagger
 =\bra{\Omega_n}P_R^\dagger .
 \label{eq:complementary-bra}
\eea
For complementary regions, writing
$M_n(t):=M_n(\bar R,R;t)$. Substitution in \eqref{eq:replica-identity} therefore gives
\bea
M_n(t)=\bra{\Omega_n}P_R^\dagger
 U_n^\dagger P_R U_n\ket{\Omega_n}=\bra{\Omega_n}
 e^{+\ii H_{n,D}t} e^{-\ii H_nt}
 \ket{\Omega_n},
 \label{eq:echo}
\eea
with the definition of the defect Hamiltonian $H_{n,D}:=P_R^\dagger H_n P_R$.  The second line follows from unitary conjugation of the exponential. This defect Hamiltonian was previously discussed by Cardy~\cite{Cardy:2011zz} in the context of measuring spatial entanglement. %The two same-oriented sewing insertions have thus become an inverse pair by using purity and complementarity. 

We now make a few remarks on the defect Hamiltonian \eqref{eq:echo}. To display its local structure, write
\begin{equation}
 H=H_{\bar R}+H_R+\sum_\mu X_\mu^{\bar R}Y_\mu^R,
 \label{eq:crossing-decomposition}
\end{equation}
where the last sum contains the interactions crossing the partition. Any crossing term can be expanded in this product form at the stated regulator. The operators $H_{\bar R}$ and $H_R$ contain terms supported entirely within the two regions.

The fixed convention \eqref{eq:cycle-action} implies
\[
 P_R^\dagger Y_{\mu,a}^R P_R=Y_{\mu,a+1}^R,
 \qquad a+n\equiv a.
\]
The sum of replica Hamiltonians internal to $R$ is invariant under this relabeling, and the operators in $\bar R$ are unchanged. It follows that
\bea
 H_{n,D}
 =\sum_{a=1}^n\left[
 H_{\bar R}^{(a)}+H_R^{(a)}
 +\sum_\mu X_{\mu,a}^{\bar R}Y_{\mu,a+1}^{R}
 \right].
 %&&V_{n,R}:=H_{n,D}-H_n=\sum_{a,\mu}X_{\mu,a}^{\bar R}
 %\left(Y_{\mu,a+1}^{R}-Y_{\mu,a}^{R}\right).
% \label{eq:defect-rewiring}
\eea
Thus a coupling that originally joined the two sides within one replica is reconnected to the adjacent replica. This is the replica-defect quench. Cardy's replica quench construction relates spatial R\'enyi moments
to overlaps between the ground states of the original and defect
Hamiltonians~\cite{Cardy:2011zz}. In our spacetime setting, the ordered
replica sewing instead expresses the STDM moments as overlaps between
states evolved from the same initial state under these two Hamiltonians.
Specifically, defining
\bea
\ket{\Omega_n(t)}=e^{-\ii H_nt}\ket{\Omega_n},
\qquad
\ket{\Omega_{D,n}(t)}=e^{-\ii H_{n,D}t}\ket{\Omega_n},
\eea
the echo in Eq.~\eqref{eq:echo} is precisely the state overlap
$\braket{\Omega_{D,n}(t)}{\Omega_n(t)}$.
This identity holds for an arbitrary pure initial state.
%without
%requiring it to be a ground state or an energy eigenstate.

As a consistency check, when the interactions across the boundary between
$R$ and $\bar R$ vanish, $H=H_{\bar R}+H_R$. The replica permutation then
leaves $H_n$ invariant, so $H_{n,D}=H_n$ and consequently
$M_n(t)=1$. This is consistent with the result in \cite{Guo:2025dtq}. This remains true for a spatially entangled pure initial state. The nontrivial temporal entanglement arises from comparing the original and reconnected couplings.

No stationarity assumption has entered this section. Next section we will impose $H\ket\psi=E_\psi\ket\psi$ to turn the echo into the return amplitude of a fixed generator and to define the associated Krylov dynamics.

\section{Krylov--R\'enyi and Krylov--von Neumann complexity}
\label{sec:krylov}

\subsection{Return amplitude in an energy eigenstate}
\label{sec:moment-return}

The replica echo derived above compares two evolutions starting
from the same state. It reduces to a return amplitude when the
initial state is an energy eigenstate,
\begin{equation}
 H\ket{\psi}=E_\psi\ket{\psi}.
 \label{eq:stationary-state}
\end{equation}
Indeed, the replicated state satisfies
$H_n\ket{\Omega_n}=nE_\psi\ket{\Omega_n}$, so its evolution under
$H_n$ contributes only an overall phase. Eq.~\eqref{eq:echo} becomes
\begin{align}
 M_n(t)
 &=\bra{\Omega_n}
   e^{+\ii H_{n,D}t}e^{-\ii H_nt}
   \ket{\Omega_n}
 \notag\\
 &=e^{-\ii nE_\psi t}
   \bra{\Omega_n}e^{+\ii H_{n,D}t}\ket{\Omega_n}
 \notag\\
 &=\bra{\Omega_n}
   e^{+\ii\mathsf H_nt}\ket{\Omega_n},
 \qquad
 \mathsf H_n:=H_{n,D}-nE_\psi\id.
 \label{eq:stationary-echo}
\end{align}
The energy shift measures the defect evolution relative to the
initial replica energy. Since $\mathsf H_n$ is self-adjoint,
we can define the normalized return amplitude
\begin{equation}
 G_n(t):=
 \bra{\Omega_n}e^{-\ii\mathsf H_nt}\ket{\Omega_n},
 \qquad
 G_n(0)=1.
 \label{eq:return-amplitude}
\end{equation}
Thus,
\begin{equation}
 M_n(t)=G_n(t)^*.
 \label{eq:survival}
\end{equation}
The complex conjugation follows from the operator ordering in the
STDM definition\footnote{For $t>0$, defining the STDM \eqref{eq:stdm-full-definition} using the time-ordered
correlator would instead give $T_{A_0B_t}^{\dagger}$, whose replica
moments are $M_n(t)^*=G_n(t)$. But in this paper we still use the anti-time-ordered
convention to keep our definitions and notation consistent with
Refs.~\cite{Milekhin:2025ycm,Guo:2025dtq}.}. This relation applies to any pure energy eigenstate.

Equation~\eqref{eq:survival} identifies a specific unitary evolution
associated with the spacetime moments: its Hamiltonian is
$\mathsf H_n$, and its initial state is $\ket{\Omega_n}$.
These are precisely the two ingredients needed to define state
spread complexity\footnote{We choose $\ket{\Omega_n}$ as the reference state because its
return amplitude under $\mathsf H_n$ is the complex conjugate of the
replica moment. This is the most direct choice for the replica
construction, although one can also define spread complexity using
other reference states with the same Hamiltonian.}. 
%We next review this construction in the present
%replica setting, keeping a finite-dimensional regulator.

\subsection{Krylov--R\'enyi complexity}
\label{sec:renyi-krylov}

Spread complexity measures how an evolving state spreads over the
ordered Krylov basis generated by its Hamiltonian and initial
state. The construction in this section closely follows the state spread
complexity framework of~\cite{Balasubramanian:2022tpr},
see also the review~\cite{Jeong:2026gdc}.
The relevant sequence of states follows directly from the
time-evolution expansion,
\begin{equation}
 e^{-\ii\mathsf H_nt}\ket{\Omega_n}
 =
 \sum_{j=0}^{\infty}
 \frac{(-\ii t)^j}{j!}\,
 \mathsf H_n^j\ket{\Omega_n}.
 \label{eq:krylov-expansion}
\end{equation}
The evolution therefore remains in the subspace
\begin{equation}
 \mathcal K_n
 =
 \operatorname{span}\left\{
 \ket{\Omega_n},
 \mathsf H_n\ket{\Omega_n},
 \mathsf H_n^2\ket{\Omega_n},\ldots
 \right\}.
 \label{eq:krylov-space}
\end{equation}
Applying Gram--Schmidt orthogonalization in this order gives
$\{\ket{K_{k,n}}\}$, with
$\ket{K_{0,n}}=\ket{\Omega_n}$.
The index $k$ labels successive independent directions generated
by repeated action of the Hamiltonian.

Self-adjointness implies that $\mathsf H_n$ is tridiagonal in this
basis:
\begin{equation}
 \mathsf H_n\ket{K_{k,n}}
 =
 b_{k+1,n}\ket{K_{k+1,n}}
 +a_{k,n}\ket{K_{k,n}}
 +b_{k,n}\ket{K_{k-1,n}},
 \label{eq:lanczos}
\end{equation}
where $a_{k,n}$ are real, $b_{0,n}=0$, and the nonzero
$b_{k,n}$ are chosen positive.
The recursion terminates when no further independent direction
is generated.

Expanding the evolved state as
\begin{equation}
 e^{-\ii\mathsf H_nt}\ket{\Omega_n}
 =
 \sum_k\varphi_{k,n}(t)\ket{K_{k,n}},
 \qquad
 \varphi_{k,n}(t)
 =
 \bra{K_{k,n}}e^{-\ii\mathsf H_nt}\ket{\Omega_n},
 \label{eq:krylov-evolution}
\end{equation}
the Schr\"odinger equation becomes
\begin{equation}
 \ii\partial_t\varphi_{k,n}
 =
 b_{k,n}\varphi_{k-1,n}
 +a_{k,n}\varphi_{k,n}
 +b_{k+1,n}\varphi_{k+1,n},
 \qquad
 \varphi_{k,n}(0)=\delta_{k0}.
 \label{eq:krylov-chain}
\end{equation}
This describes propagation along a chain whose sites are the
Krylov basis vectors. Unitarity gives
$\sum_k|\varphi_{k,n}(t)|^2=1$.
We define the \textit{Krylov--R\'enyi complexity} associated with the region $R$
as the mean position along this chain,
\bea
 \mathcal C_{n}(t):=
 \sum_k k\,|\varphi_{k,n}(t)|^2.
 \label{eq:complexity}
\eea
It is the standard state spread complexity of the pair
$(\mathsf H_n,\ket{\Omega_n})$.
The replica number $n$ labels the family of evolutions selected
by the STDM construction.

The return amplitude is the amplitude at the initial Krylov site:
\begin{equation}
 \varphi_{0,n}(t)=G_n(t)=M_n(t)^*.
 \label{eq:krylov-return}
\end{equation}
Its full time dependence also supplies the data needed to construct
the remaining Krylov dynamics. In particular,
\begin{equation}
 \mu_{j,n}
 :=
 \bra{\Omega_n}\mathsf H_n^j\ket{\Omega_n}
 =
 \left.\ii^j\partial_t^jG_n(t)\right|_{t=0}.
 \label{eq:return-moments}
\end{equation}
These moments determine the inner products used in the
orthogonalization,
\begin{equation}
 \left\langle
 \mathsf H_n^r\Omega_n
 \,\middle|\,
 \mathsf H_n^s\Omega_n
 \right\rangle
 =
 \mu_{r+s,n}.
 \label{eq:krylov-gram}
\end{equation}
They consequently determine the Lanczos coefficients and the
chain evolution in Eq.~\eqref{eq:krylov-chain}.
The complete return function
therefore determines the spread complexity.
This reconstruction uses the time dependence of $M_n(t)$;
its value at a single time only fixes the initial-site amplitude.

Timelike Rényi entropy and Krylov–Rényi complexity characterize different aspects of the same replica-defect evolution. In the setting considered above,
\bea
e^{(1-n)S_n^{\rm time}(t)}
=M_n(t)=\varphi_{0,n}(t)^*
\eea
Comparing this expression with the definition of complexity in
\eqref{eq:complexity}, we see that, at a fixed time, the entropy
encodes the magnitude and phase of the return amplitude, whereas
the complexity measures the average position of the evolving
state along the Krylov chain. They therefore provide complementary dynamical characterizations: the return amplitude at a single time does not, in general, determine how the probability away from the initial state is distributed over the remaining Krylov sites. However, this distinction does not imply independent dynamical information. The complete time dependence of the complex moment$M_n(t)$, obtainable from the timelike Rényi entropy, determines the Lanczos coefficients and hence the entire complexity evolution. Thus, information inaccessible from the entropy at a single time can nevertheless be reconstructed from its full time dependence.

\subsection{Equivalent formulation with the original Hamiltonian}
\label{sec:permutation-complexity}
The use of the defect Hamiltonian $H_{n,D}$ in the above definition raises
the question of how this complexity characterizes the dynamics
of the original system.
In this subsection, we show that the Krylov--R\'enyi complexity
is exactly equal to the spread complexity of the
seed state $P_R\ket{\Omega_n}$ evolving under
the original replicated Hamiltonian $H_n$. 

To establish this equivalence, define
\begin{equation}
 \ket{\Phi_{n,R}}:=P_R\ket{\Omega_n}.
 \label{eq:permutation-state}
\end{equation}
Since $P_R$ is unitary, this state is normalized.
The definition of the defect Hamiltonian gives
\begin{equation}
 \mathsf H_n
 =
 P_R^\dagger
 (H_n-nE_\psi\id)
 P_R,
 \label{eq:generator-equivalence}
\end{equation}
and hence
\begin{equation}
 G_n(t)
 =
 \bra{\Phi_{n,R}}
 e^{-\ii(H_n-nE_\psi\id)t}
 \ket{\Phi_{n,R}}.
 \label{eq:permutation-return}
\end{equation}

To compare the spread complexities, we must also compare their
Krylov bases. For every nonnegative integer $j$,
\begin{equation}
 (H_n-nE_\psi\id)^j\ket{\Phi_{n,R}}
 =
 P_R\mathsf H_n^j\ket{\Omega_n}.
 \label{eq:krylov-sequence-map}
\end{equation}
Because the permutation preserves all inner products,
Gram--Schmidt orthogonalization of the two sequences gives
\begin{equation}
 \ket{\widetilde K_{k,n}}
 =
 P_R\ket{K_{k,n}},
 \label{eq:krylov-basis-map}
\end{equation}
with consistent phase conventions.

The energy shift does not change this ordered Krylov basis.
Indeed, the binomial expansion expresses
$(H_n-nE_\psi\id)^j\ket{\Phi_{n,R}}$ as
$H_n^j\ket{\Phi_{n,R}}$ plus lower powers of $H_n$.
Those lower-power contributions are removed during
orthogonalization. Thus $\{\ket{\widetilde K_{k,n}}\}$ is also
the Krylov basis generated directly by
$(H_n,\ket{\Phi_{n,R}})$.

The corresponding evolution amplitudes satisfy
\begin{align}
 \widetilde\varphi_{k,n}(t)
 &:=
 \bra{\widetilde K_{k,n}}
 e^{-\ii H_nt}\ket{\Phi_{n,R}}
 \notag\\
 &=
 \bra{K_{k,n}}
 P_R^\dagger e^{-\ii H_nt}
 P_R\ket{\Omega_n}
 \notag\\
 &=
 e^{-\ii nE_\psi t}\varphi_{k,n}(t).
 \label{eq:krylov-amplitude-map}
\end{align}
All Krylov probabilities therefore agree, and
\begin{equation}
 \sum_k k\,|\widetilde\varphi_{k,n}(t)|^2
 =
 \sum_k k\,|\varphi_{k,n}(t)|^2
 =
 \mathcal C_{n}(t).
 \label{eq:complexity-equivalence}
\end{equation}
Thus the R\'enyi complexity defined through the STDM moments is
therefore exactly the spread complexity of the permutation
excitation $P_R\ket{\Omega_n}$ evolving under the original
replicated Hamiltonian $H_n$.

\subsection{Calculation by spectral measure method}
\label{sec:spectral-calculation}

We now describe how to compute the Krylov--R\'enyi complexity
directly from the return amplitude using its spectral measure.
We use the standard spectral representation of Krylov dynamics
in terms of orthogonal polynomials~\cite{Muck:2022xfc,Caputa:2025ozd}
to compute the Krylov--R\'enyi complexity from the return amplitude
determined by the STDM moments.
This formulation will also provide the starting point for
the von Neumann limit discussed below.

The full time dependence of the return amplitude determines the
remaining Krylov amplitudes. To make this reconstruction explicit,
we use the spectral measure of $\mathsf H_n$ in the initial state
$\ket{\Omega_n}$,
\begin{equation}
 d\nu_n(E):=\bra{\Omega_n}d\Pi_n(E)\ket{\Omega_n},
 \qquad
 G_n(t)=\int_{\mathbb R}e^{-\ii Et}\,d\nu_n(E),
 \label{eq:spectral-measure}
\end{equation}
where $d\Pi_n(E)$ is the spectral resolution of $\mathsf H_n$.
The measure is positive and normalized, $\int_{\mathbb R}d\nu_n(E)=1$\footnote{For a ground-state initial condition, $\mathsf H_n\geq0$,
and the spectral measure is supported on $[0,\infty)$.
For a general energy eigenstate, its support may also include
negative energies, so we retain integration over $\mathbb R$.}.

Orthogonalizing $1,E,E^2,\ldots$ with respect to $d\nu_n$
gives real orthonormal polynomials $p_{k,n}(E)$, with
$p_{0,n}=1$ and positive leading coefficients.
This is the spectral representation of the same Gram--Schmidt
construction that defines the Krylov basis:
\begin{equation}
 \ket{K_{k,n}}=p_{k,n}(\mathsf H_n)\ket{\Omega_n}.
 \label{eq:spectral-krylov-basis}
\end{equation}
Consequently,
\begin{align}
 \varphi_{k,n}(t)
 &=\int_{\mathbb R}p_{k,n}(E)e^{-\ii Et}\,d\nu_n(E),
 \notag\\
 \mathcal C_{n}(t)
 &=\sum_k k\left|
   \int_{\mathbb R}p_{k,n}(E)e^{-\ii Et}\,d\nu_n(E)
   \right|^2.
 \label{eq:spectral-complexity}
\end{align}
Thus the complete return function $G_n(t)=M_n(t)^*$ fixes
$d\nu_n$, which in turn fixes the orthonormal polynomials
and the complexity. This is a spectral formulation of the
standard reconstruction of spread complexity from the return
amplitude~\cite{Balasubramanian:2022tpr}. The derivation and the equivalent reconstruction from spectral moments are given in Appendix~\ref{app:spectral-reconstruction}.

\subsection{Field-theory realization and regularization}
\label{sec:qft-regularization}

In a field theory, the regional permutation $P_R$ is represented
by a replica permutation defect. For a finite interval $R=[u,v]$ in two spacetime dimensions,
the permutation-excited seed is represented by branch-point
twist fields:
\begin{equation}
 \ket{\Phi_{n,R}}=P_R\ket{\Omega_n}
 \propto
 \sigma_n(u)\widetilde{\sigma}_n(v)\ket{\Omega_n}.
\end{equation}
The two twist fields implement opposite cyclic permutations,
and the proportionality includes a cutoff-dependent
normalization~\cite{Cardy:2011zz,Cardy:2007mb}.
In higher dimensions, the corresponding replica defect is
supported on $\partial R$.

Sharp defect insertions can produce divergent energy moments
in the continuum limit~\cite{Cardy:2011zz}.
We suppress these ultraviolet contributions by evolving the
seed through a Euclidean time $\delta/2$ and then normalizing:
\begin{equation}
 \ket{\Phi_{n,R}^{[\delta]}}
 :=\frac{
 e^{-\frac{\delta}{2}(H_n-nE_\psi\id)}
 \ket{\Phi_{n,R}}
 }{\sqrt{F_n(\delta)}},
 \qquad \delta>0.
 \label{eq:regulated-permutation-state}
\end{equation}
Here $F_n(\delta)$ is the norm squared before normalization.
For a general Euclidean separation $\tau>0$, it is defined by
\begin{equation}
 F_n(\tau)
 :=\bra{\Phi_{n,R}}
 e^{-\tau(H_n-nE_\psi\id)}
 \ket{\Phi_{n,R}},
 \label{eq:euclidean-twist-amplitude}
\end{equation}
with all operator expressions initially understood at a common
microscopic cutoff.
Thus $\delta$ is the total Euclidean separation between the
bra and ket insertions. Equivalently, $F_n(\tau)$ is the
Euclidean defect correlator, which becomes a twist four-point
function for a finite interval in two dimensions.

The corresponding return amplitude is
\begin{align}
 G_n^{[\delta]}(t)
 &:=\bra{\Phi_{n,R}^{[\delta]}}
 e^{-\ii(H_n-nE_\psi\id)t}
 \ket{\Phi_{n,R}^{[\delta]}}
 =\frac{F_n(\delta+\ii t)}{F_n(\delta)},
 \notag\\
 G_n^{[\delta]}(t)^*
 &=\frac{M_n(t+\ii\delta)}{M_n(\ii\delta)}.
 \label{eq:regulated-return-amplitude}
\end{align}
The second line follows from $F_n(\tau)=M_n(\ii\tau)$
with the same cutoff and analytic-continuation prescription.
Any common multiplicative normalization of the twist insertions
cancels in these ratios, and $G_n^{[\delta]}(0)=1$.

Using the original definition of the timelike R\'enyi entropy,
we obtain
\begin{equation}
 \frac{\operatorname{Log}\!\left[G_n^{[\delta]}(t)^*\right]}{1-n}
 =
 S_n^{\rm time}(t+\ii\delta)
 -S_n^{\rm time}(\ii\delta),
 \label{eq:regulated-return-renyi}
\end{equation}
with logarithmic branches chosen consistently from $t=0$.
The imaginary time shift implements the regularization,
while the subtraction follows from normalizing the regulated
seed. Thus the return amplitude encodes the regulated entropy
change relative to $t=0$, without changing the original
definition of $S_n^{\rm time}$.

For field-theory calculations, we can also apply the spectral method of
Sec.~\ref{sec:spectral-calculation} to this normalized return
amplitude. Equivalently, the spectral measure becomes
\begin{equation}
 d\nu_n^{[\delta]}(E)
 =\frac{e^{-\delta E}\,d\nu_n(E)}{F_n(\delta)}.
 \label{eq:regulated-spectral-measure}
\end{equation}
The Krylov polynomials and amplitudes are constructed using
this regulated measure, and the resulting probabilities give
the complexity through Eq.~\eqref{eq:complexity}.

The superscript $[\delta]$ makes the regulator dependence explicit.
For notational simplicity, we suppress this superscript in subsequent
field-theory calculations, while all quantities remain understood
at fixed $\delta>0$. As the examples below illustrate, the resulting
complexity depends explicitly on the regulator.

\subsection{Krylov--von Neumann complexity}
\label{sec:von-neumann-complexity}

At $n=1$, the replica permutation becomes the identity and
the seed reduces to the original energy eigenstate.
With the energy shift used above, $G_1(t)=1$ and
$\mathcal C_{1}(t)=0$.
The nontrivial replica limit therefore concerns the leading
variation of the complexity near $n=1$.
For a specified continuation to real $n>1$ admitting a
normalized positive spectral measure, we define
\begin{equation}
 \mathcal C_{\mathrm{vN}}(t)
 :=\lim_{\lambda\to 0}
 \frac{\mathcal C_{1+\lambda}(t)}{\lambda},
 \quad \text{with} \ \lambda=n-1,
 \label{eq:von-neumann-complexity}
\end{equation}
whenever the limit exists.
The region and preparation are held fixed.
%In field theory, this includes the regulator $\delta>0$,
%with the superscript $[\delta]$ suppressed as in
%Sec.~\ref{sec:qft-regularization}.

The input for this calculation is the first replica
variation of the normalized return amplitude,
\begin{equation}
 G_{1+\lambda}(t)=1-\lambda\Psi(t)+o(\lambda),
 \qquad \Psi(0)=0.
 \label{eq:vn-return-expansion}
\end{equation}
For the regulated field-theory preparation,
Eq.~\eqref{eq:regulated-return-amplitude} relates this
function directly to the timelike von Neumann entropy:
\begin{equation}
 \Psi(t)^*
 =S^{\rm time}(t+\ii\delta)-S^{\rm time}(\ii\delta),
 \qquad
 S^{\rm time}(z)
 :=-\left.\partial_n\log M_n(z)\right|_{n=1}.
 \label{eq:vn-entropy-input}
\end{equation}
The same analytic-continuation prescription is used
throughout.

To evaluate Eq.~\eqref{eq:von-neumann-complexity},
we need the $O(\sqrt{\lambda})$ contribution to each
nonzero-site amplitude, since
\begin{equation}
 \frac{\mathcal C_{1+\lambda,R}(t)}{\lambda}
 =\sum_{k\ge1}k
 \left|\frac{\varphi_{k,1+\lambda}(t)}
 {\sqrt{\lambda}}\right|^2.
 \label{eq:vn-amplitude-target}
\end{equation}
Appendix~\ref{app:vn-spectral-limit} extracts these
contributions from the finite-replica spectral
representation under a regular replica expansion.
At each fixed degree before termination, the leading
nonconstant part of the original Krylov polynomial
defines $q_j$ through
\[
 \sqrt{\lambda}\,
 \big[p_{j+1,1+\lambda}(E)-p_{j+1,1+\lambda}(0)\big]
 \longrightarrow E q_j(E).
\]
The original orthogonality relation then fixes the
measure that normalizes these limiting polynomials:
\begin{equation}
 d\Lambda(E)
 :=\lim_{\lambda\to 0}
 \frac{E^2}{\lambda}\,d\nu_{1+\lambda}(E),
 \qquad
 \Psi''(t)=\int_{\mathbb R}e^{-\ii Et}\,d\Lambda(E).
 \label{eq:vn-curvature-measure}
\end{equation}
The second relation follows by taking two time derivatives
of the return function.
The spectral and moment limits used here are specified
in Appendix~\ref{app:vn-spectral-limit}. Thus $d\Lambda$ can be obtained from the Fourier transform
of $\Psi''$.
Its moments, which provide the input for the polynomial
recursion, are
\begin{equation}
 m_{j+2}:=\int_{\mathbb R}E^j\,d\Lambda(E)
 =\left.\ii^j\partial_t^{j+2}\Psi(t)\right|_{t=0},
 \qquad j\ge0.
 \label{eq:vn-curvature-moments}
\end{equation}
For $m_2=\Psi''(0)>0$, we can show the limiting polynomials satisfy
\begin{equation}
 \int_{\mathbb R}q_j(E)q_l(E)\,d\Lambda(E)=\delta_{jl},
 \qquad q_0(E)=m_2^{-1/2},
 \label{eq:vn-curvature-polynomials}
\end{equation}
with positive leading coefficients.
They can therefore be constructed directly from $d\Lambda$
or its moments.

Their three-term recurrence is inherited from the
original Lanczos construction:
\begin{equation}
 E q_j(E)
 =\beta_{j+1}q_{j+1}(E)
  +\alpha_jq_j(E)+\beta_jq_{j-1}(E),
 \qquad q_{-1}=0,\quad\beta_0=0,
 \label{eq:vn-three-term}
\end{equation}
where
\[
 \alpha_j=\lim_{\lambda\to 0}a_{j+1,1+\lambda},
 \qquad
 \beta_{j+1}=\lim_{\lambda\to 0}b_{j+2,1+\lambda}.
\]
The index shift reflects the correspondence between
$q_j$ and the original polynomial $p_{j+1,n}$.
The original first link satisfies
$b_{1,1+\lambda}^2=\lambda m_2+o(\lambda)$ and vanishes
in the limit, while $\alpha_j$ and $\beta_{j+1}$
describe the remaining chain starting at site $k=1$.
The explicit recursion for computing these coefficients
and polynomials is given in
Eq.~\eqref{eq:vn-polynomial-recursion} of
Appendix~\ref{app:vn-spectral-limit}.

Using this polynomial limit in the spectral amplitudes
gives
\bea
 &&\varphi_{k,1+\lambda}(t)
 =\sqrt{\lambda}\,V_k(t)+o(\sqrt{\lambda}),
 \nn\\
 &&V_k(t)=\int_{\mathbb R}q_{k-1}(E)
 \frac{e^{-\ii Et}-1}{E}\,d\Lambda(E),
 \qquad k\ge1.
 \label{eq:vn-leading-amplitude}
\eea
The subtraction of $1$ follows from the orthogonality
of $p_{k,n}$ to the constant polynomial and implements
$\varphi_{k,n}(0)=0$.
The quotient takes its continuous value $-\ii t$ at
$E=0$, retaining any spectral contribution there.

Assuming that the weighted sum converges and can be
interchanged with the replica limit, we obtain
\begin{equation}
 \mathcal C_{\mathrm{vN}}(t)
 =\sum_{j\ge0}(j+1)
 \left|
 \int_{\mathbb R}q_j(E)
 \frac{e^{-\ii Et}-1}{E}\,d\Lambda(E)
 \right|^2.
 \label{eq:vn-spectral-result}
\end{equation}
If the limiting polynomial sequence terminates,
the sum runs only over the existing polynomials.
Completeness provides the consistency check
$\sum_{k\ge1}|V_k(t)|^2=2\operatorname{Re}\Psi(t)$,
matching the leading probability of leaving the
initial Krylov site.
For $m_2=0$, positivity gives $d\Lambda=0$ and the
leading escape probability vanishes, which leads to
$\mathcal C_{\mathrm{vN}}(t)=0$.

Consequently, the full timelike von Neumann
entropy determines the input $\Psi(t)$, whose second
time derivative fixes the measure, the limiting
Lanczos coefficients, and the leading amplitudes\footnote{As in the replica calculation of the von Neumann entropy, we analytically continue the normalized return amplitude $G_n(t)$
from integer replica number to real $n>1$ before taking the
limit $n\to1$. We assume that this continuation admits a
positive spectral measure for $n$ sufficiently close to one.
Whether such a continuation exists and is uniquely selected in
general remains an open question. For a given continuation,
the first-order replica response $\Psi(t)$ determines
$\mathcal C_{\mathrm{vN}}(t)$ uniquely through $\Psi''(t)$
under the convergence assumptions of
Appendix~\ref{app:vn-spectral-limit}.}.
Only the first replica variation is required.
The derivation and consistency checks are presented
in Appendix~\ref{app:vn-spectral-limit}.

\subsection{Short-time expansion}
\label{sec:short-time-expansion}

The short-time expansion makes explicit which features of
timelike entanglement determine the initial growth of
complexity. Throughout, the preparation is fixed, in field
theory, the regulator $\delta>0$ is also held fixed.

The spectral representation gives
$\varphi_{k,n}(t)=O(t^k)$, and each Krylov probability is
even in $t$. Probability conservation therefore implies
\begin{equation}
 \mathcal C_n(t)
 =1-|G_n(t)|^2+|\varphi_{2,n}(t)|^2+O(t^6).
 \label{eq:short-time-second-site}
\end{equation}
Thus the probability of leaving the initial state determines
the quadratic growth, while the second Krylov site supplies
the first additional contribution.

Define the energy cumulants by
\[
 \kappa_{r,n}
 :=\left.\ii^r\partial_t^r\log G_n(t)\right|_{t=0},
 \qquad \log G_n(0)=0.
\]
Expanding the two contributions in
Eq.~\eqref{eq:short-time-second-site} gives
\begin{equation}
 \mathcal C_n(t)
 =\kappa_{2,n}t^2+
 \left(
 \frac{\kappa_{4,n}}{6}
 -\frac{\kappa_{3,n}^2}{4\kappa_{2,n}}
 \right)t^4+O(t^6),
 \label{eq:short-time-cumulants}
\end{equation}
for $\kappa_{2,n}>0$.
The derivation and the required spectral regularity are
given in Appendix~\ref{app:short-time-expansion}. 
If the energy variance $\kappa_{2,n}$ vanishes, the seed
has a single energy and $\mathcal C_n(t)=0$. An equivalent short-time expansion in terms of Lanczos coefficients was given in~\cite{Aguilar-Gutierrez:2023nyk,Caputa:2025ozd}.

To express this result in terms of timelike entanglement,
we write
\begin{equation}
 R_n(t)+\ii I_n(t)
 =\frac{\log[G_n(t)^*]}{1-n},
 \label{eq:short-time-entropy-data}
\end{equation}
where $R_n(t)$ and $I_n(t)$ denote the real and imaginary
parts of the timelike R\'enyi entropy, respectively\footnote{
In field theory, $G_n(t)$ denotes the normalized regulated
return amplitude, with the superscript $[\delta]$ suppressed.
Accordingly, $R_n(t)+\ii I_n(t)
=S_n^{\rm time}(t+\ii\delta)-S_n^{\rm time}(\ii\delta)$
at fixed $\delta>0$, as in
Eq.~\eqref{eq:regulated-return-renyi}.}.
The cumulant expansion then yields
\begin{equation}
 \begin{aligned}
 \frac{\mathcal C_n(t)}{n-1}
 =
 R_n''(0)t^2-\left[
 \frac{R_n^{(4)}(0)}{6}
 +\frac{[I_n'''(0)]^2}{4R_n''(0)}
 \right]t^4+O(t^6).
 \end{aligned}
 \label{eq:short-time-renyi}
\end{equation}
The curvature of the real part fixes the initial growth.
The imaginary part first contributes through its cubic
time derivative at order $t^4$; its linear term produces
only an overall phase and does not affect complexity.
The first correction therefore already probes information
beyond the return probability.

For the von Neumann complexity, expanding the limiting
amplitudes in Eq.~\eqref{eq:vn-leading-amplitude} gives
\begin{equation}
 \mathcal C_{\mathrm{vN}}(t)
 =m_2t^2+
 \left(
 \frac{m_4}{6}-\frac{m_3^2}{4m_2}
 \right)t^4+O(t^6),
 \label{eq:short-time-vn}
\end{equation}
where
$m_r=\int_{\mathbb R}E^{r-2}\,d\Lambda(E)$ for $r\ge2$
and $m_2>0$.
Writing $R(t)+\ii I(t)=\Psi(t)^*$, these coefficients obey
$m_2=R''(0)$, $m_3=I'''(0)$, and $m_4=-R^{(4)}(0)$.
Thus the same entropy derivatives determine the initial
growth in the replica limit.
Under the assumptions of
Sec.~\ref{sec:von-neumann-complexity}, the case $m_2=0$
instead gives $\mathcal C_{\mathrm{vN}}(t)=0$.

\section{Examples in quantum field theories}
\label{sec:examples}

\subsection{A half-line in 2-dimensional CFT}
\label{sec:half-line-cft}

We first consider the half-line example in 2-dimensional CFT.
Take the vacuum on the infinite line,
with central charge $c$. We choose $R=(0,\infty)$. There is a single spatial cut at $x=0$.
The corresponding replica defect ends on a branch-point
twist field and its conjugate, so the field-theory input
is a vacuum two-point function~\cite{Cardy:2007mb,Cardy:2011zz}:
\bea
 F_n(\tau)=\big\langle \widetilde{\sigma}_n(0,\tau)\mathcal \sigma_n(0,0)\big\rangle
 \propto\tau^{-2\Delta_n},
 \label{eq:half-line-twist-kernel}
\eea
where $\Delta_n=\frac{c}{12}\left(n-\frac1n\right)$ is the
scaling dimension of the twist fields. Using the preparation of
Sec.~\ref{sec:qft-regularization}, we obtain
\begin{equation}
 G_n(t)
 =\frac{F_n(\delta+\ii t)}{F_n(\delta)}
 =(1+\ii u)^{-2\Delta_n},
 \label{eq:half-line-return}
\end{equation}
where $u:=\frac{t}{\delta}$, $\delta>0$ is the regulator as shown in Sec.~\ref{sec:qft-regularization}.
%and the branch is fixed by
%continuation from $t=0$.

The normalized STDM moment is $G_n(t)^*$.
Consequently, according to~\eqref{eq:short-time-entropy-data} the real and imaginary parts are given by
\begin{equation}
 R_n(t)=\frac{\Delta_n}{n-1}\log(1+u^2),
 \qquad
 I_n(t)=-\frac{2\Delta_n}{n-1}\arctan u.
 \label{eq:half-line-entropy}
\end{equation}
In particular,  we can see that both vanish at $t=0$.

To compute the complexity, we now extract the spectral
measure from Eq.~\eqref{eq:half-line-return}.
The Gamma integral gives
\begin{equation}
 G_n(t)=\int_0^\infty e^{-\ii Et}\,d\nu_n(E),
 \qquad
 d\nu_n(E)
 =\frac{\delta^{2\Delta_n}}{\Gamma(2\Delta_n)}
 E^{2\Delta_n-1}e^{-\delta E}\,dE.
 \label{eq:half-line-measure}
\end{equation}
This is a normalized positive measure of the excitation
energy above the replicated vacuum.

According to Eq.~\eqref{eq:spectral-complexity}, computing
the Krylov amplitudes requires the orthonormal polynomials
of the measure $d\nu_n$.
We therefore first identify these polynomials for the
Gamma measure in Eq.~\eqref{eq:half-line-measure}.
With $x=\delta E$, this measure becomes
\[
 d\nu_n(E)
 =\frac{x^{2\Delta_n-1}e^{-x}}{\Gamma(2\Delta_n)}\,dx.
\]
The generalized Laguerre polynomials obey
\[
 \int_0^\infty x^{a-1}e^{-x}
 L_k^{(a-1)}(x)L_l^{(a-1)}(x)\,dx
 =\frac{\Gamma(a+k)}{k!}\delta_{kl},
 \qquad a>0.
\]
Setting $a=2\Delta_n$ and accounting for the normalization
of $d\nu_n$, we obtain
\begin{equation}
 p_{k,n}(E)
 =(-1)^k
 \sqrt{\frac{k!}{(2\Delta_n)_k}}\,
 L_k^{(2\Delta_n-1)}(\delta E),
 \label{eq:half-line-polynomials}
\end{equation}
where $(a)_k=\Gamma(a+k)/\Gamma(a)$.
These polynomials satisfy the orthonormality condition
in Eq.~\eqref{eq:spectral-orthogonality}.
The factor $(-1)^k$ makes their leading coefficients
positive, fixing the sign convention.
Orthonormality with respect to $d\nu_n$, together with the
positive leading-coefficient convention, uniquely identifies
these as the Krylov polynomials; they therefore satisfy the
Lanczos recurrence in Eq.~\eqref{eq:spectral-three-term}.

We next substitute Eqs.~\eqref{eq:half-line-measure}
and \eqref{eq:half-line-polynomials} into the amplitude
formula in Eq.~\eqref{eq:spectral-complexity}.
Changing variables to $x=\delta E$ gives
\[
 \varphi_{k,n}(t)
 =\frac{(-1)^k}{\Gamma(2\Delta_n)}
 \sqrt{\frac{k!}{(2\Delta_n)_k}}
 \int_0^\infty
 x^{2\Delta_n-1}e^{-(1+\ii u)x}
 L_k^{(2\Delta_n-1)}(x)\,dx.
\]
This integral is evaluated using the Laguerre Laplace
transform,
\begin{equation}
 \frac{1}{\Gamma(a)}
 \int_0^\infty x^{a-1}e^{-s x}
 L_k^{(a-1)}(x)\,dx
 =\frac{(a)_k}{k!}\frac{(s-1)^k}{s^{a+k}},
 \qquad a>0,\quad \operatorname{Re}s>0.
 \label{eq:half-line-laguerre-transform}
\end{equation}
Taking $a=2\Delta_n$ and $s=1+\ii u$, we find
\begin{equation}
 \varphi_{k,n}(t)
 =\sqrt{\frac{(2\Delta_n)_k}{k!}}\,
 \frac{(-\ii u)^k}{(1+\ii u)^{2\Delta_n+k}}.
 \label{eq:half-line-amplitudes}
\end{equation}
At $k=0$, this reproduces the return amplitude
in Eq.~\eqref{eq:half-line-return}. Now we can obtain the Krylov-R\'enyi complexity
\bea
\mathcal C_n(t)=\frac{c}{6}\left(n-\frac1n\right)\frac{t^2}{\delta^2}.
\label{eq:half-line-complexity}
\eea

%Writing $z=u^2/(1+u^2)$, the Krylov probabilities take
%the negative binomial form
%\[
% |\varphi_{k,n}(t)|^2
% =(1-z)^{2\Delta_n}\frac{(2\Delta_n)_k}{k!}z^k.
%\]
%The generating function
%$\sum_{k\ge0}(2\Delta_n)_kz^k/k!=(1-z)^{-2\Delta_n}$
%ensures normalization and gives the mean Krylov depth:
%\begin{equation}
% \begin{aligned}
% \mathcal C_n(t)
% &=(1-z)^{2\Delta_n}
%   z\frac{d}{dz}(1-z)^{-2\Delta_n}
% \\
 %&=2\Delta_n\frac{z}{1-z}
% =\frac{c}{6}\left(n-\frac1n\right)
%   \frac{t^2}{\delta^2}.
% \end{aligned}
% \label{eq:half-line-complexity}
%\end{equation}
%For this continuum vacuum preparation, the quadratic law
%holds at all real times.
%As a check, the cumulants
%$\kappa_{r,n}=2\Delta_n(r-1)!/\delta^r$
%make the quartic coefficient in
%Eq.~\eqref{eq:short-time-cumulants} vanish.

The result agrees with the known spread complexity of a
regulated local primary excitation of dimension $\Delta_n$,
with one-sided Euclidean preparation time
$\delta/2$~\cite{Caputa:2024sux}.
In the present construction, the excitation is the replica
twist field, whose dimension is fixed by $n$ and the
central charge.

The Gamma measure remains positive for every real $n>1$,
providing a spectral continuation for the replica limit.
Taking this limit in the exact expression
\eqref{eq:half-line-complexity} gives
\bea
 \mathcal C_{\mathrm{vN}}(t)
 =\lim_{n\to 1}
   \frac{\mathcal C_n(t)}{n-1}
 =\frac{c}{3}\frac{t^2}{\delta^2}.
 \label{eq:half-line-vn-complexity}
\eea
%The first-order data entering
%Sec.~\ref{sec:von-neumann-complexity} are correspondingly
%\[
% \Psi(t)=\frac c3\log(1+\ii t/\delta),
% \qquad
% d\Lambda(E)=\frac c3 E e^{-\delta E}\,dE,
%\]
%which satisfy
%$\Psi''(t)=\int_0^\infty e^{-\ii Et}\,d\Lambda(E)$.
%Both complexities depend explicitly on the regulator.

%Finally, eliminating $t/\delta$ between
%Eqs.~\eqref{eq:half-line-entropy}
%and \eqref{eq:half-line-complexity} gives
%\begin{equation}
% R_n(t)
% =\frac{\Delta_n}{n-1}
% \log\!\left(
% 1+\frac{\mathcal C_n(t)}{2\Delta_n}
 %\right).
% \label{eq:half-line-entropy-complexity}
%\end{equation}
%Thus the real part of the subtracted entropy grows
%logarithmically at late times, while complexity grows
%quadratically.
%In this example, the real part at a single time already
%determines the complexity because the vacuum two-point
%function fixes the entire functional form.
%For a finite interval, both spatial endpoints contribute,
%and the relevant input becomes a twist four-point function,
%as we discuss next.

\subsection{A finite interval in the massless Dirac theory}
\label{sec:dirac-finite-interval}

We now consider a finite interval $R=[0,\ell]$ in the vacuum
of the massless complex Dirac theory on the infinite line,
with central charge $c=1$ for a single
species~\cite{Casini:2005rm,Headrick:2012fk}.
The two subsystems are $\bar R$ at time zero and $R$ at
time $t$. We use the normalized, regulated state defined
in Sec.~\ref{sec:qft-regularization}, with fixed $\delta>0$.
The return amplitude is determined by a twist four-point
function, allowing us to study how the additional length
scale $\ell$ affects Krylov spreading.

\subsubsection{From the spatial correlator to the return amplitude}

The required Euclidean correlator is the one that determines
the spatial two-interval replica moment. Writing $z_{ab}=z_a-z_b$
and $\gamma_n=2\Delta_n=c(n-n^{-1})/6$, it is
\begin{equation}
 \big\langle\sigma_n(z_1)\widetilde{\sigma}_n(z_2)
 \sigma_n(z_3)\widetilde{\sigma}_n(z_4)\big\rangle
 =A_n\left|
 \frac{z_{13}z_{24}}{z_{12}z_{14}z_{23}z_{34}}
 \right|^{\gamma_n},
 \label{eq:dirac-four-point}
\end{equation}
where $A_n$ contains the endpoint normalization~\cite{Casini:2005rm}.
In the return amplitude, the lower cut inserts $P_R$ and the
upper cut inserts $P_R^\dagger$. Their endpoints therefore have
opposite spatial order: $(z_1,z_2,z_3,z_4)
=(0,\ell,\ell+\ii\tau,\ii\tau)$, with $\tau>0$.
The numerator in Eq.~\eqref{eq:dirac-four-point} has modulus
$\ell^2+\tau^2$, while the denominator has modulus
$\ell^2\tau^2$. Hence
\begin{equation}
 F_n(\tau)=A_n
 \left(\frac{\ell^2+\tau^2}{\ell^2\tau^2}\right)^{\gamma_n}.
 \label{eq:dirac-euclidean}
\end{equation}

Using the normalized continuation in
Eq.~\eqref{eq:regulated-return-amplitude}, we obtain
\begin{equation}
 G_n(t)=\frac{F_n(\delta+\ii t)}{F_n(\delta)}
 =\left[
 \frac{1+\ell^2/(\delta+\ii t)^2}{1+\ell^2/\delta^2}
 \right]^{\gamma_n}.
 \label{eq:dirac-return}
\end{equation}
The power is defined by continuing the logarithm from $t=0$.
More explicitly, introduce
\begin{equation}
 f_\ell(s)=\log(s+\ii\ell)+\log(s-\ii\ell)-2\log s,
 \qquad \operatorname{Re}s>0,
 \label{eq:dirac-log-kernel}
\end{equation}
with each logarithm on its principal branch. Then
$\log G_n(t)=\gamma_n[f_\ell(\delta+\ii t)-f_\ell(\delta)]$.
%The normalized STDM moment is $G_n(t)^*$, with the ordering
%specified in Sec.~\ref{sec:sewing}.
For $\ell\to\infty$ at fixed $\delta,t$, this amplitude becomes
$(1+\ii t/\delta)^{-2\gamma_n}$, the product of the two
half-line amplitudes in Eq.~\eqref{eq:half-line-return}.

\subsubsection{The spectral measure and the complexity}

To apply Sec.~\ref{sec:spectral-calculation}, we need the energy
measure whose Fourier transform is Eq.~\eqref{eq:dirac-return}.
The integral representation of $f_\ell$ gives
\bea
 \log G_n(t)=\gamma_n\int_0^\infty(e^{-\ii Et}-1)j(E)\,dE,\quad \text{with}\ 
 j(E)=\frac{2e^{-\delta E}}{E}[1-\cos(\ell E)].
 \label{eq:dirac-jump}
\eea
Exponentiating and expanding in $\gamma_n$ yields
\bea
 d\nu_n(E)=e^{-\gamma_nJ}
 \left[\delta(E)dE+\sum_{p=1}^\infty
 \frac{\gamma_n^p}{p!}\,j^{*p}(E)\,dE\right],
 \label{eq:dirac-measure}
\eea
where $J=\int_0^\infty j(E)\,dE=\log(1+\ell^2/\delta^2)$.
Here $j^{*p}$ denotes the $p$-fold convolution of $j$.
The initial condition at $p=1$ is $j^{*1}(E)=j(E)$.
Higher convolutions are generated recursively by
\bea
 j^{*(p+1)}(E)
 =\int_0^E j^{*p}(E')j(E-E')\,dE',
 \qquad p\ge1.
\eea
The term $\delta(E)dE$ denotes a delta function at $E=0$,
and its coefficient is the zero-energy spectral weight.
%Since $j\ge0$, Eq.~\eqref{eq:dirac-measure} is positive and
%normalized for every real $n>1$. It therefore also specifies
%the positive replica continuation used below.
The derivation of the measure is given in
Appendix~\ref{app:dirac-spectrum}.

The energy moments of this measure determine $p_{k,n}$ and the
Lanczos coefficients through Eq.~\eqref{eq:spectral-recursion}.
Their overlaps can then be evaluated directly from $G_n$ using
Eq.~\eqref{eq:spectral-differential}:
\begin{equation}
 \varphi_{k,n}(t)=p_{k,n}(\ii\partial_t)G_n(t),
 \qquad
 \mathcal C_n(t)=\sum_{k\ge1}k\,|\varphi_{k,n}(t)|^2.
 \label{eq:dirac-renyi-series}
\end{equation}
Appendix~\ref{app:dirac-spectrum} gives the cumulants and the
moment recursion needed to implement this calculation.
In this example, we do not obtain a simple closed-form expression
for the complexity. Once the spectral measure is known, the
polynomials $p_{k,n}$ can be constructed recursively and used
to evaluate Eq.~\eqref{eq:dirac-renyi-series}.
This expansion over Krylov levels is valid at any fixed time
without requiring $t/\delta\ll1$.
At large times, evaluating the sum generally requires more Krylov levels.

For the von Neumann complexity, we apply the first-order construction
of Sec.~\ref{sec:von-neumann-complexity}. We have
\begin{equation}
 \Psi(t)=\frac c3
 [f_\ell(\delta)-f_\ell(\delta+\ii t)].
 \label{eq:dirac-psi}
\end{equation}
The measure required by Eq.~\eqref{eq:vn-curvature-measure}
is obtained by differentiating this expression twice:
\begin{equation}
 \Psi''(t)=\int_0^\infty e^{-\ii Et}\,d\Lambda(E),
 \qquad
 d\Lambda(E)=\frac{2c}{3}E e^{-\delta E}
 [1-\cos(\ell E)]\,dE.
 \label{eq:dirac-curvature}
\end{equation}
Thus the interval length enters the limiting Krylov problem
through the factor $1-\cos(\ell E)$.
Constructing the polynomials $q_j$ of this measure with
Eq.~\eqref{eq:vn-polynomial-recursion} gives
\begin{equation}
 V_{j+1}(t)=\int_0^\infty q_j(E)
 \frac{e^{-\ii Et}-1}{E}\,d\Lambda(E),
 \qquad
 \mathcal C_{\mathrm{vN}}(t)
 =\sum_{j\ge0}(j+1)|V_{j+1}(t)|^2.
 \label{eq:dirac-vn-series}
\end{equation}
Appendix~\ref{app:dirac-vn-calculation} evaluates each integral
as a finite sum. 
%Appendix~\ref{app:dirac-convergence} shows that
%the resulting weighted series converges and equals the replica
%limit at every fixed real $t$ and $\ell,\delta>0$.

\subsubsection{The effect of the interval length}

Although we do not obtain a simple closed-form expression in
this example, the short- and long-time limits reveal how the
subsystem size affects the complexity. We focus on the
Krylov--von Neumann complexity $\mathcal C_{\mathrm{vN}}$.

The short-time expansion follows directly from the results
of Sec.~\ref{sec:short-time-expansion}. Substituting the first three moments of $d\Lambda$ into
Eq.~\eqref{eq:short-time-vn} gives
\bea
 \mathcal C_{\mathrm{vN}}(t)=\frac{2c}{3}
 \left[
 \frac{r^2(r^2+3)}{(1+r^2)^2}u^2
 +\frac{r^2(r^2-6)}{(1+r^2)^2(r^2+3)}u^4
 \right]+O(u^6),
 \label{eq:dirac-short}
\eea
where $u=t/\delta$ and $r=\ell/\delta>0$.
%Through order $u^4$, the finite-$n$ expansion is obtained
%by replacing $2c/3$ with $2\gamma_n$.
For well-separated endpoints, $r\to\infty$, the initial
quadratic coefficient approaches twice the half-line value
and the quartic correction vanishes.

For long time limit, at fixed $\ell,\delta>0$, Eq.~\eqref{eq:dirac-return} gives
\begin{equation}
 G_n(t)=G_n(\infty)
 \left[1-\frac{\gamma_n\ell^2}{t^2}+O(t^{-3})\right],
 \qquad
 G_n(\infty)=(1+r^2)^{-\gamma_n}.
 \label{eq:dirac-return-late}
\end{equation}
The normalized timelike R\'enyi quantity
$\log[G_n(t)^*]/(1-n)$ therefore approaches
$\gamma_n\log(1+r^2)/(n-1)$.
Its von Neumann limit is the real plateau
\begin{equation}
 \Psi_\infty:=\lim_{t\to\infty}\Psi(t)
 =\frac c3\log(1+r^2).
 \label{eq:dirac-entropy-plateau}
\end{equation}
This plateau fixes the total leading escape weight,
$\sum_{k\ge1}|V_k(t)|^2\to2\Psi_\infty$, whereas the
complexity also measures the Krylov depth reached by that weight.
The large-degree analysis in Appendix~\ref{app:dirac-long-time}
relates the oscillatory factor $e^{-\ii Et}$ to degrees
$k\simeq t^2E/\delta$, giving
\begin{equation}
 \mathcal C_{\mathrm{vN}}(t)
 =\frac{2c}{3}\frac{r^2}{1+r^2}u^2+o(u^2),
 \qquad u\to+\infty.
 \label{eq:dirac-long}
\end{equation}
Here the replica limit is taken first, with $\ell$ and $\delta$
fixed. For $\ell\gg\delta$, the coefficient of the late-time $t^2$ growth
is $\frac{2c}{3}\bigl(\delta^{-2}-\ell^{-2}
\bigr)$. The leading term is set by the UV
cutoff, while the first finite-interval correction,
$-2c/(3\ell^2)$, is independent of $\delta$. We find this behavior for a finite interval in the massless Dirac theory. Whether a similar separation between the cutoff-dependent leading term
and the cutoff-independent finite-interval correction holds in other
quantum field theories remains an open question. 

The coefficient also admits a simple expression in terms
of the regulator dependence of the entropy plateau:
\begin{equation}
 \lim_{t\to\infty}\frac{\mathcal C_{\mathrm{vN}}(t)}{t^2}
 =-\frac1\delta
   \left.\partial_\delta\Psi_\infty\right|_{\ell,c}.
 \label{eq:dirac-plateau-growth}
\end{equation}
This relation follows from the spectral structure of the present
Dirac example. It doesn't hold for general cases. Generally, a constant return amplitude alone does not imply quadratic growth. For every finite $r>0$, the late-time quadratic
coefficient is smaller than the initial one by the ratio
$(1+r^2)/(3+r^2)$.

\subsection{A planar cut in 3-dimensional CFT}
\label{sec:planar-strip-cft}

\subsubsection{From strip entropy to the first replica variation}

We next apply the method of Sec.~\ref{sec:von-neumann-complexity}
to the vacuum of a $(2+1)$-dimensional CFT. We consider $R=\{x>0\}$ on the $(x,y)$ plane and regulate the cut on the $y$-direction with length of $L$. In this example, entanglement of timelike strip determines the first replica variation needed
for $\mathcal C_{\mathrm{vN}}$.

Let $s>0$ denote the Euclidean time separation between
the permutation insertions $P_R$ and $P_R^\dagger$,
placed at $\tau=-s/2$ and $\tau=s/2$, respectively.
Each insertion introduces a replica cut across $R=\{x>0\}$.
Relabeling the replicas in the region
$x>0$, $-s/2<\tau<s/2$ replaces these two cuts by a
single replica cut at $x=0$, $|\tau|<s/2$.
This cut forms a strip of width $s$ along the $\tau$
direction and regulated length $L$ along the $y$ direction.

Denote the replica moments of the strip by
$F_{n,L}(s)$.
Taking the replica derivative at $n=1$, we obtain
\begin{equation}
 -\left.\partial_n\log
 \frac{F_{n,L}(s)}{F_{n,L}(\delta)}\right|_{n=1}
 =
 S_{\mathrm{strip},L}(s)-S_{\mathrm{strip},L}(\delta),
 \label{eq:strip-entropy-difference}
\end{equation}
where $S_{\mathrm{strip},L}(s)$ is the entanglement entropy
of the strip.

Conformal invariance fixes the
width-dependent extensive part of the strip entropy to
$-\sigma_1 L/s$, where $\sigma_1>0$ is a dimensionless,
theory-dependent coefficient\footnote{This scaling form is expected for a spatial strip
in the vacuum of a 3-dimensional CFT in the planar
limit~\cite{Casini:2006hu}. Euclidean rotational invariance
allows the same result to be applied to the strip extending
along the $\tau$ direction considered here.}.
The UV area term is independent of the width and cancels
in Eq.~\eqref{eq:strip-entropy-difference}. Consequently,
\begin{equation}
 S_{\mathrm{strip},L}(s)-S_{\mathrm{strip},L}(\delta)
 =
 L\sigma_1\left(\frac1\delta-\frac1s\right).
 \label{eq:strip-coefficient}
\end{equation}

To compute $\mathcal C_{\mathrm{vN}}$, we first determine
the replica variation $\Psi(t)$ introduced in
Eq.~\eqref{eq:vn-return-expansion}.
The normalized return amplitude is
$G_n(t)=F_{n,L}(\delta+\ii t)/F_{n,L}(\delta)$.
Continuing Eq.~\eqref{eq:strip-entropy-difference} to
$s=\delta+\ii t$ and using Eq.~\eqref{eq:strip-coefficient},
we obtain
\begin{align}
 \Psi(t)
 &=-\left.\partial_n\log G_n(t)\right|_{n=1}
 \notag\\
 &=S_{\mathrm{strip},L}(\delta+\ii t)
   -S_{\mathrm{strip},L}(\delta)
 \notag\\
 &=\sigma_1L
 \left(\frac1\delta-\frac1{\delta+\ii t}\right).
 \label{eq:strip-psi}
\end{align}
We take the replica limit at fixed $L$, followed by
$L\to\infty$ at fixed $\delta$ and $t$.
The calculation below assumes the spectral conditions of
Sec.~\ref{sec:von-neumann-complexity} and that this planar
limit can be taken through the analytic continuation,
spectral integrals, and weighted Krylov sum.

\subsubsection{Computing the Krylov--von Neumann complexity}

We now compute $\mathcal C_{\mathrm{vN}}(t)$ from the
first replica variation in Eq.~\eqref{eq:strip-psi},
following the method of
Sec.~\ref{sec:von-neumann-complexity}.
%Throughout this calculation, we retain the leading
%extensive contribution in $L$.
According to Eq.~\eqref{eq:vn-curvature-measure},
$\Psi''(t)$ determines the measure $d\Lambda$ through
its Fourier representation. For the strip, we have
\begin{equation}
 \Psi''(t)
 =\frac{2\sigma_1L}{(\delta+\ii t)^3}
 =\int_0^\infty e^{-\ii Et}\,d\Lambda(E),
 \qquad
 d\Lambda(E)=\sigma_1L E^2e^{-\delta E}\,dE.
 \label{eq:strip-curvature}
\end{equation}
This measure is positive, with total weight
$m_2=2\sigma_1L/\delta^3$.

To construct the polynomials entering the leading
amplitudes, set $x=\delta E$. The measure is then
proportional to the Laguerre weight $x^2e^{-x}dx$.
Using
\[
 \int_0^\infty x^2e^{-x}
 L_j^{(2)}(x)L_l^{(2)}(x)\,dx
 =(j+1)(j+2)\delta_{jl},
\]
we obtain
\begin{equation}
 q_j(E)=(-1)^j
 \sqrt{\frac{\delta^3}{\sigma_1L(j+1)(j+2)}}\,
 L_j^{(2)}(\delta E).
 \label{eq:strip-polynomials}
\end{equation}
These polynomials satisfy
$\int q_jq_l\,d\Lambda=\delta_{jl}$ and have positive
leading coefficients, so they are precisely the
polynomials required in
Eq.~\eqref{eq:vn-leading-amplitude}.
Their recurrence coefficients in
Eq.~\eqref{eq:vn-three-term} are
\[
 \alpha_j=\frac{2j+3}{\delta},
 \qquad
 \beta_{j+1}=\frac{\sqrt{(j+1)(j+3)}}{\delta}.
\]

Substituting Eq.~\eqref{eq:strip-polynomials} into
Eq.~\eqref{eq:vn-leading-amplitude}, and writing
$u=t/\delta$, gives
\begin{equation}
 V_k(t)=(-1)^{k-1}
 \sqrt{\frac{\sigma_1L}{\delta k(k+1)}}
 \int_0^\infty xe^{-x}L_{k-1}^{(2)}(x)
                  \big(e^{-\ii ux}-1\big)\,dx,
 \qquad k\ge1.
 \label{eq:strip-amplitude-integral}
\end{equation}
%To evaluate this integral, use
%$L_{k-1}^{(2)}=\sum_{a=0}^{k-1}L_a^{(1)}$
%and the Laguerre Laplace transform from
%Sec.~\ref{sec:half-line-cft}.
%With $w=\ii u/(1+\ii u)$, the exponential term gives
%\begin{align*}
% \int_0^\infty xe^{-(1+\ii u)x}L_{k-1}^{(2)}(x)\,dx
% &=(1-w)^2\sum_{a=0}^{k-1}(a+1)w^a\\
% &=1-w^k[(k+1)-kw].
%\end{align*}
%Its value at $u=0$ is $1$. Subtracting this value as
%required in Eq.~\eqref{eq:strip-amplitude-integral}
%therefore yields
With some calculations, we have
\begin{equation}
 V_k(t)=(-1)^k
 \sqrt{\frac{\sigma_1L}{\delta k(k+1)}}\,
 w^k[(k+1)-kw],
 \label{eq:strip-amplitudes}
\end{equation}
where $w=\ii u/(1+\ii u)$.  As a consistency check, the unweighted sum satisfies
\begin{equation}
 \sum_{k\ge1}|V_k(t)|^2
 =
 \frac{2\sigma_1 L}{\delta}\frac{u^2}{1+u^2}
 =
 2\operatorname{Re}\Psi(t).
 \label{eq:strip-escape-check}
\end{equation}
%The complexity follows from
%$\mathcal C_{\mathrm{vN}}(t)=\sum_{k\ge1}k|V_k(t)|^2$,
%as in Eq.~\eqref{eq:vn-spectral-result}.
%To perform the sum, set
%$z=|w|^2=\operatorname{Re}w=u^2/(1+u^2)$.
%Then
%\[
% |(k+1)-kw|^2=(k+1)^2(1-z)+z,
%\]
%and the geometric and logarithmic series give
%\begin{align}
 %\lim_{L\to\infty}\frac{\mathcal C_{\mathrm{vN}}(t)}{L}
% &=
% \frac{\sigma_1}{\delta}
% \sum_{k\ge1}z^k
% \left[(k+1)(1-z)+\frac{z}{k+1}\right]
% \notag\\
% &=
% \frac{\sigma_1}{\delta}
% \left[\frac{z}{1-z}-\log(1-z)\right]
% \notag\\
% &=
% \frac{\sigma_1}{\delta}
 %\left[u^2+\log(1+u^2)\right].
% \label{eq:strip-response}
%\end{align}
%The series converges for every fixed $t$, since $z<1$.
The Krylov-von Neumann complexity follows from the definition $\mathcal C_{\mathrm{vN}}(t)=\sum_{k\ge1}k|V_k(t)|^2$, the result is
\bea
C_{\mathrm{vN}}(t)=\frac{\sigma_1 L}{\delta}\left[u^2+\log(1+u^2)\right].
\label{eq:strip-response}
\eea

The short-time expansion of
Eq.~\eqref{eq:strip-response} is
\bea
\mathcal C_{\mathrm{vN}}(t)
 =
 \frac{\sigma_1 L}{\delta}
 \left[2u^2-\frac{u^4}{2}+O(u^6)\right].
\eea
One could check the result agrees with Eq.~\eqref{eq:short-time-vn}.
At late times,
\bea
 \mathcal C_{\mathrm{vN}}(t)
 =
 \frac{\sigma_1L}{\delta}
 \left[u^2+2\log|u|+O(u^{-2})\right].
\eea
Thus the strip example gives a closed expression for the
complexity density from the first replica variation,
with the theory dependence entering through $\sigma_1$. A notable feature of this higher-dimensional example is the subleading logarithmic correction to the quadratic
growth of the complexity at late times.

\section{Summary and discussion}\label{Sec_discussion}

We have established a common dynamical framework for timelike entanglement and Krylov complexity based on the spacetime density matrix. For stationary pure states and complementary regions at different times, spacetime replica moments determine the return amplitude of a replica-defect quench. The associated Krylov–Rényi complexity is the standard state spread complexity of this evolution. We also define Krylov-von Neumann compelxity as the limit $n\to 1$.  The field-theory examples demonstrate how the two quantities distinguish different aspects of the same dynamics. In particular, for a finite interval in massless Dirac theory at fixed ultraviolet regulator, timelike entropy approaches a constant while Krylov–von Neumann complexity grows quadratically at late times, showing that entropy saturation can coexist with continued Krylov spreading. We also show the framework can be applied to higher dimensional examples. In the vacuum field-theory examples studied here, Krylov--von Neumann complexity continues to grow at late times, even when the timelike entanglement entropy approaches a constant. Studies of holographic
complexity in black hole backgrounds suggest that nonperturbative effects can give rise to a late-time plateau~\cite{Miyaji:2025yvm,Miyaji:2025jxy}. It would be interesting to apply our replica construction in these
settings to determine whether Krylov--von Neumann complexity also develops such a plateau.

Our results also provide a method for computing Krylov complexity from timelike entanglement. Under the conditions established in this work, the full time dependence of the complex timelike von Neumann entropy supplies the data needed to reconstruct Krylov–von Neumann complexity. Existing analytical results for timelike entanglement can therefore furnish further examples for investigating Krylov dynamics in quantum field theory and holographic settings.

The framework also suggests a numerical approach, which we have not pursued here. In Gaussian lattice models, timelike Rényi and von Neumann entropies can be evaluated efficiently using Gaussian methods~\cite{Milekhin:2025ycm,Wang:2026hxj}. Their time dependence could then serve as input for reconstructing Krylov–Rényi and Krylov–von Neumann complexity, respectively. Developing stable numerical algorithms for this reconstruction and testing their convergence would extend the range of systems accessible to the present framework.

The present construction focuses on stationary pure states and complementary retained regions, for which the spacetime replica moments are identified with the complex conjugate of a return amplitude, as expressed in Eq.~\eqref{eq:survival}. For a nonstationary pure state, the moments still admit the two-Hamiltonian echo representation in Eq.~\eqref{eq:echo}, involving the overlap between states evolved under the original and defect Hamiltonians. How to associate a natural Krylov complexity with this more general evolution while retaining its connection to timelike entanglement remains an interesting question. The approach to Krylov complexity based on density-matrix operators developed in~\cite{Caputa:2024vrn} may be useful in extending
our construction to more general initial states. The defect-quench construction of complexity also admits a straightforward extension to stationary mixed states, including thermal states. We will present this extension and its relation to spacetime replica observables elsewhere.

Finally, our construction suggests a route toward a holographic description of Krylov complexity. Both state spread complexity and operator Krylov complexity require a specified seed state or operator, and their physical interpretation depends on this choice. In the present framework, the seed is fixed by the regional replica construction once the background state, subsystem geometry, and regulator are specified, without introducing an independently chosen probe operator. Like entanglement entropy, this construction is defined nonperturbatively and retains dynamical information about the underlying theory. These features motivate the search for universal structures in the resulting complexity. Although holographic relations for spread complexity have been established in particular settings~\cite{Caputa:2024sux}, a general geometric prescription remains to be developed. The Ryu–Takayanagi formula for spatial entanglement provides a well-established precedent in semiclassical AdS/CFT~\cite{Ryu:2006bv}. Our connection between timelike entanglement and Krylov complexity raises the possibility that the replica construction may similarly lead to a universal bulk prescription for Krylov–Rényi and Krylov–von Neumann complexity. Exploring this possibility and identifying the bulk quantity that captures the associated Krylov spreading are directions we plan to pursue.

The bulk interpretation of spread complexity is particularly clear in
low-dimensional holographic models. In double-scaled SYK, for example,
the spread complexity of thermofield double states has been related to
bulk geodesic length, including the length of the two-sided wormhole in
the JT gravity limit~\cite{Rabinovici:2023yex,Heller:2024ldz,Aguilar-Gutierrez:2025pqp,Fu:2025kkh}.
It would be interesting to study whether the Krylov complexity defined
by our replica construction also admits a geometric interpretation in
these models.

\section*{Acknowledgments}

I thank Fabio Ori, Juan Pedraza, Shan-Ming Ruan and Tadashi Takayanagi for helpful discussions and valuable comments on the draft.
I am grateful
to the organizers of the workshop \textit{Holographic Universe 2026},
held at the Yukawa Institute for Theoretical Physics (YITP) in
May 2026. The talks on complexity at this workshop inspired me
to explore the connection between spacetime density matrices
and Krylov complexity. I used ChatGPT (GPT-6) to help work out parts of the calculations,
particularly the replica limit and the late-time behavior of the
finite-interval example, and to improve the clarity of the manuscript.
I take full responsibility for the results and their interpretation.
This work was supported by the National Natural Science Foundation of China under Grant No.~12675068 and the Hubei Provincial Natural Science Foundation of China under Grant No.~2025AFB557. I also thank the support from the Seventh Young Faculty Development Program of Huazhong University of Science and Technology.

\appendix
\section{Proof of the replica sewing identity}
\label{app:sewing-details}

We derive \eqref{eq:replica-identity} from the matrix elements
\eqref{eq:stdm-components}. For $\rho=\ket\psi\bra\psi$,
expanding the matrix product gives
\begin{align}
 M_n(A,B;t)
 &=\sum_{\{a_j,b_j\}}
   \prod_{j=1}^n T_{a_jb_j,a_{j+1}b_{j+1}}
 \notag\\
 &=\sum_{\{a_j,b_j\}}
   \prod_{j=1}^n
   \bra\psi
   (\ket{a_{j+1}}\bra{a_j})_A
   U^\dagger
   (\ket{b_{j+1}}\bra{b_j})_B
   U\ket\psi ,
 \label{eq:sewing-contractions}
\end{align}
where $a_{n+1}=a_1$ and $b_{n+1}=b_1$. As in the main text,
each subsystem operator acts as the identity on its complement.

The product of expectation values is an expectation value in
$\ket{\Omega_n}=\ket\psi^{\otimes n}$. Using
$U_n=U^{\otimes n}$ and tensor-product multiplication, we obtain
\begin{align}
 M_n(A,B;t)
 =\bra{\Omega_n}
 &\left[
   \sum_{\{a_j\}}\bigotimes_{j=1}^n
   (\ket{a_{j+1}}\bra{a_j})_A
  \right] U_n^\dagger
 \notag\\[-1mm]
 {}\times&
 \left[
   \sum_{\{b_j\}}\bigotimes_{j=1}^n
   (\ket{b_{j+1}}\bra{b_j})_B
  \right] U_n
 \ket{\Omega_n}.
 \label{eq:sewing-regrouped}
\end{align}
This regrouping preserves the operator order within every replica.

For any subsystem $R$, the operator
\[
 \sum_{\{r_j\}}\bigotimes_{j=1}^n
 (\ket{r_{j+1}}\bra{r_j})_R,
 \qquad r_{n+1}=r_1,
\]
maps $\ket{r_1,\ldots,r_n}$ to
$\ket{r_2,\ldots,r_n,r_1}$ and acts trivially on the complementary
factors. It is therefore precisely $P_R$ with the convention
\eqref{eq:cycle-action}. Substituting this identification into
\eqref{eq:sewing-regrouped} yields
\begin{equation}
 M_n(A,B;t)
 =\bra{\Omega_n}P_A U_n^\dagger
                    P_B U_n\ket{\Omega_n}
 =\Tr\!\left[
   \rho_n P_A U_n^\dagger P_B U_n
  \right],
\end{equation}
which proves \eqref{eq:replica-identity}. Both regional cycles
have the same orientation because both sets of indices are
contracted as $j\to j+1$. No stationarity or locality assumption
is required.

\section{Spectral methods for Krylov complexity}
\label{app:spectral-reconstruction}

\subsection{Spectral representation and reconstruction at fixed replica number}
\label{app:spectral-fixed-n}

We show how the return amplitude determines the Krylov evolution
without requiring an explicit construction of the replica Hilbert
space. Throughout the finite-dimensional discussion, $n$ is fixed,
$\ket{\Omega_n}$ is normalized, and
$\mathsf H_n=H_{n,D}-nE_\psi\id$ is the self-adjoint generator
introduced in Eq.~\eqref{eq:stationary-echo}.

The spectral theorem gives
\begin{equation}
 \bra{\Omega_n}f(\mathsf H_n)\ket{\Omega_n}
 =\int_{\mathbb R}f(E)\,d\nu_n(E).
 \label{eq:spectral-functional}
\end{equation}
In finite dimension, this measure takes the form
\begin{equation}
 d\nu_n(E)
 =\sum_{\alpha}w_{\alpha,n}
   \delta(E-E_{\alpha,n})\,dE,
 \qquad
 w_{\alpha,n}
 =\bra{\Omega_n}\Pi_{\alpha,n}\ket{\Omega_n},
 \label{eq:spectral-discrete}
\end{equation}
where $E_{\alpha,n}$ are the distinct eigenvalues of
$\mathsf H_n$ with nonzero weight, and $\Pi_{\alpha,n}$
are their spectral projectors.
The weights satisfy $w_{\alpha,n}>0$ and
$\sum_\alpha w_{\alpha,n}=1$.
Taking $f(E)=e^{-\ii Et}$ gives
Eq.~\eqref{eq:spectral-measure}. Conversely, knowledge of
the complex function $G_n(t)$ for all real $t$ uniquely
determines $d\nu_n$ by uniqueness of the Fourier transform
of finite measures.

To relate this measure to the Krylov basis, equip polynomials
with the inner product
\begin{equation}
 \langle f,g\rangle_{\nu_n}
 :=\int_{\mathbb R}f(E)^*g(E)\,d\nu_n(E)
 =\bra{\Omega_n}
   f(\mathsf H_n)^\dagger g(\mathsf H_n)
   \ket{\Omega_n}.
 \label{eq:spectral-isometry}
\end{equation}
The polynomial inner product is therefore identical to that
of the corresponding Krylov vectors. Gram--Schmidt applied
to $1,E,E^2,\ldots$ gives precisely
Eq.~\eqref{eq:spectral-krylov-basis}, with
\begin{equation}
 \int_{\mathbb R}p_{j,n}(E)p_{k,n}(E)\,d\nu_n(E)
 =\delta_{jk}.
 \label{eq:spectral-orthogonality}
\end{equation}
Positive leading coefficients implement the convention
$b_{k,n}>0$ for every nonzero Lanczos coefficient.

For completeness, the polynomials and Lanczos coefficients can
be generated directly from the measure. Starting from
$p_{-1,n}=0$, $p_{0,n}=1$, and $b_{0,n}=0$, set
\begin{align}
 a_{k,n}
 &=\int_{\mathbb R}E\,p_{k,n}(E)^2\,d\nu_n(E),
 \notag\\
 r_{k+1,n}(E)
 &=(E-a_{k,n})p_{k,n}(E)-b_{k,n}p_{k-1,n}(E),
 \notag\\
 b_{k+1,n}^2
 &=\int_{\mathbb R}r_{k+1,n}(E)^2\,d\nu_n(E),
 \qquad
 p_{k+1,n}(E)=\frac{r_{k+1,n}(E)}{b_{k+1,n}}.
 \label{eq:spectral-recursion}
\end{align}
The last expression applies when $b_{k+1,n}>0$; a zero
norm terminates the chain. To see why only three adjacent
polynomials appear, note that for $j\leq k-2$,
\[
 \langle p_{j,n},E p_{k,n}\rangle_{\nu_n}
 =\langle E p_{j,n},p_{k,n}\rangle_{\nu_n}=0,
\]
because $E p_{j,n}$ has degree at most $k-1$.
The coefficient of $p_{k-1,n}$ equals $b_{k,n}$ by the
same symmetry. Hence, before termination,
\begin{equation}
 E p_{k,n}
 =b_{k+1,n}p_{k+1,n}
  +a_{k,n}p_{k,n}
  +b_{k,n}p_{k-1,n},
 \label{eq:spectral-three-term}
\end{equation}
which maps to Eq.~\eqref{eq:lanczos} under
Eq.~\eqref{eq:spectral-isometry}.

Taking the overlap of
$p_{k,n}(\mathsf H_n)\ket{\Omega_n}$
with the evolved state gives
Eq.~\eqref{eq:spectral-complexity}.
Equivalently, the amplitudes can be computed directly from
the return function:
\begin{equation}
 \varphi_{k,n}(t)
 =p_{k,n}(\ii\partial_t)G_n(t),
 \label{eq:spectral-differential}
\end{equation}
since $\ii\partial_t e^{-\ii Et}=E e^{-\ii Et}$.
Orthogonality gives
$\varphi_{k,n}(0)=\delta_{k0}$.
Moreover, in finite dimension the polynomials span all functions
on the support of $d\nu_n$, so Parseval's identity yields
\begin{equation}
 \sum_k|\varphi_{k,n}(t)|^2
 =\int_{\mathbb R}|e^{-\ii Et}|^2\,d\nu_n(E)=1.
 \label{eq:spectral-normalization}
\end{equation}

An explicit Fourier inversion is not necessary if the
derivatives of $G_n(t)$ at the origin are available.
They determine the spectral moments
\begin{equation}
 \mu_{j,n}
 :=\int_{\mathbb R}E^j\,d\nu_n(E)
 =\bra{\Omega_n}\mathsf H_n^j\ket{\Omega_n}
 =\left.\ii^j\partial_t^jG_n(t)\right|_{t=0}.
 \label{eq:return-moments}
\end{equation}
These moments determine the polynomial inner products,
\begin{equation}
 \langle E^r,E^s\rangle_{\nu_n}
 =\mu_{r+s,n}
 =\left\langle
   \mathsf H_n^r\Omega_n
   \,\middle|\,
   \mathsf H_n^s\Omega_n
  \right\rangle,
 \label{eq:krylov-gram}
\end{equation}
and therefore all steps of Eq.~\eqref{eq:spectral-recursion}.
For example,
\begin{equation}
 a_{0,n}=\mu_{1,n},
 \qquad
 b_{1,n}^2=\mu_{2,n}-\mu_{1,n}^2,
 \qquad
 p_{1,n}(E)=\frac{E-\mu_{1,n}}{b_{1,n}}.
 \label{eq:spectral-first-step}
\end{equation}
When $b_{1,n}>0$, this gives
\begin{equation}
 \varphi_{1,n}(t)
 =\frac{\ii\partial_tG_n(t)-\mu_{1,n}G_n(t)}{b_{1,n}}.
 \label{eq:spectral-first-amplitude}
\end{equation}
If $b_{1,n}=0$, the initial state is an eigenstate of
$\mathsf H_n$; the chain terminates at $k=0$ and
$\mathcal C_{n,R}(t)=0$.
Thus either the spectral measure or its moments provide a
complete reconstruction at finite dimension.

For an infinite-dimensional system, the spectral representation
of $G_n(t)$ and its uniqueness remain valid for a normalized
state and a self-adjoint generator.
The polynomial construction additionally requires finite
moments of all orders. If the polynomials are dense in
$L^2(d\nu_n)$, the Krylov expansion is complete and
Eq.~\eqref{eq:spectral-normalization} continues to hold;
a finite complexity further requires convergence of
$\sum_k k|\varphi_{k,n}(t)|^2$.
Reconstruction from moments alone also requires care:
without a determinacy condition, the moment sequence need not
uniquely determine the measure. This ambiguity is absent
when the full return function is given.

\subsection{Spectral construction of the von Neumann limit}
\label{app:vn-spectral-limit}

We derive the von Neumann complexity from the finite-replica
spectral representation established in
Appendix~\ref{app:spectral-fixed-n}. 
%Throughout, $t$, the region, and the positive regulator, when present, are held fixed, and
%$\lambda=n-1\downarrow0$ along the specified positive spectral continuation. 
For sufficiently small $\lambda>0$, we assume
that $d\nu_{1+\lambda}$ is a normalized positive measure with
a complete polynomial Krylov representation.

\subsubsection{Surviving terms in the replica limit}
The starting point is
\begin{align}
 \varphi_{k,1+\lambda}(t)
 &=\int_{\mathbb R}p_{k,1+\lambda}(E)e^{-\ii Et}
                   \,d\nu_{1+\lambda}(E),
 \notag\\
 \frac{\mathcal C_{1+\lambda,R}(t)}{\lambda}
 &=\sum_{k\ge1}k
   \left|\frac{\varphi_{k,1+\lambda}(t)}
                    {\sqrt{\lambda}}\right|^2.
 \label{eq:vn-appendix-target}
\end{align}
Our immediate objective is therefore to determine the
coefficient of $\sqrt{\lambda}$ in each nonzero-site
amplitude, keeping its full time dependence.
Completeness and the return expansion
$G_{1+\lambda}(t)=1-\lambda\Psi(t)+o(\lambda)$ give
\begin{equation}
 \sum_{k\ge1}|\varphi_{k,1+\lambda}(t)|^2
 =1-|G_{1+\lambda}(t)|^2
 =2\lambda\operatorname{Re}\Psi(t)+o(\lambda).
 \label{eq:vn-total-escape}
\end{equation}
Positivity bounds each amplitude by $O(\sqrt{\lambda})$,
but does not by itself establish the existence of its
rescaled limit. Every fixed Krylov level can contribute at
this order. After determining these limits, we must also
justify their weighted sum in
Eq.~\eqref{eq:vn-appendix-target}.

Both the measure and the normalized polynomials in the
amplitude depend on $\lambda$. We first determine the
polynomial scaling from the original orthogonality and
normalization conditions. The required input is the
expansion of the energy moments:
\begin{equation}
 \mu_{r,1+\lambda}
 :=\int_{\mathbb R}E^r\,d\nu_{1+\lambda}(E)
 =\lambda m_r+o(\lambda),\qquad r\ge1.
 \label{eq:vn-moment-limits}
\end{equation}
We assume that all these moments exist and that the expansion
holds at every fixed order. Here $m_r$ is the first-order
coefficient of the $r$th moment.
This scaling can be checked explicitly for the first few
energy moments using the moment--derivative relation in
Eq.~\eqref{eq:return-moments}.
We first consider $m_2>0$.

At a fixed degree $k\ge1$, write the original Krylov polynomial
in its general form,
\begin{equation}
 p_{k,1+\lambda}(E)
 =c_{k0}(\lambda)
  +\sum_{r=1}^{k}c_{kr}(\lambda)E^r.
 \label{eq:vn-original-coefficient-expansion}
\end{equation}
Its orthogonality to $p_0=1$ fixes the constant coefficient:
\begin{equation}
 c_{k0}(\lambda)
 =-\sum_{r=1}^{k}
   c_{kr}(\lambda)\mu_{r,1+\lambda}.
 \label{eq:vn-constant-coefficient}
\end{equation}
Substituting this expression into the normalization condition
gives
\begin{equation}
 1=\sum_{r,s=1}^{k}c_{kr}(\lambda)c_{ks}(\lambda)
 \left[
  \mu_{r+s,1+\lambda}
  -\mu_{r,1+\lambda}\mu_{s,1+\lambda}
 \right].
 \label{eq:vn-coefficient-normalization}
\end{equation}
The moment expansion implies
\[
 \mu_{r+s,1+\lambda}
 -\mu_{r,1+\lambda}\mu_{s,1+\lambda}
 =\lambda m_{r+s}+o(\lambda),
\]
because the product of the two moments is $O(\lambda^2)$. Thus Eq.~\eqref{eq:vn-coefficient-normalization} would require $c_{kr}(\lambda)=O(\lambda^{-1/2})$ for $1\le r\le k$.

%We carry out the following argument at each fixed degree
%for which the limiting matrix
%$(m_{r+s})_{r,s=1}^{k}$ is positive definite.
%This is the range before the limiting orthogonalization
%terminates; degenerate cases are treated separately below.
%At such a degree, the quadratic form in
%Eq.~\eqref{eq:vn-coefficient-normalization}, divided by
%$\lambda$, has a strictly positive lower bound.
%Normalization consequently requires
%\[
% c_{kr}(\lambda)=O(\lambda^{-1/2}),
% \qquad 1\le r\le k.
%\]
As a result Eq.~\eqref{eq:vn-constant-coefficient} then gives
\begin{equation}
 p_{k,1+\lambda}(0)=c_{k0}(\lambda)
 =O(\sqrt{\lambda}).
 \label{eq:vn-constant-scaling}
\end{equation}
%Thus the constant term of
%$\sqrt{\lambda}\,p_{k,1+\lambda}$ vanishes in the limit.
Thus we have
\begin{equation}
 \sqrt{\lambda}\,p_{k,1+\lambda}(E)
 =\sqrt{\lambda}\,
 \big[p_{k,1+\lambda}(E)-p_{k,1+\lambda}(0)\big]
 +O(\lambda).
 \label{eq:vn-polynomial-limit}
\end{equation}
The leading nonconstant part therefore defines a
degree-$(k-1)$ polynomial $q_{k-1}$ through
\begin{equation}
 q_{k-1}(E)
 :=\lim_{\lambda\to0}\frac{\sqrt{\lambda}}{E}
 \big[p_{k,1+\lambda}(E)-p_{k,1+\lambda}(0)\big].
 \label{eq:vn-rescaled-polynomial}
\end{equation}
We next determine the normalization of $q_j$ from the
orthogonality of the original Krylov polynomials:
\[
 \int_{\mathbb R}
 p_{j+1,1+\lambda}(E)p_{l+1,1+\lambda}(E)
 \,d\nu_{1+\lambda}(E)=\delta_{jl}.
\]
Using Eq.~\eqref{eq:vn-rescaled-polynomial}, the leading order gives
\[
 \int_{\mathbb R}q_j(E)q_l(E)
 \frac{E^2}{\lambda}\,d\nu_{1+\lambda}(E)
 =\delta_{jl}+o(1).
\]
The constant terms contribute only $O(\lambda)$ to these
inner products, since
$p_{k,1+\lambda}(0)=O(\sqrt{\lambda})$.
Thus the orthogonality relation identifies the measure
associated with $q_j$:
\begin{equation}
 d\Lambda_\lambda(E)
 :=\frac{E^2}{\lambda}\,d\nu_{1+\lambda}(E).
 \label{eq:vn-prelimit-measure}
\end{equation}
%The two factors of $E$ come from the two polynomials
%in the inner product, while $1/\lambda$ comes from
%their normalization.

This measure is positive, and its total weight remains
finite in the limit $\lambda\to 0$:
\[
 \int_{\mathbb R}d\Lambda_\lambda(E)
 =\frac{\mu_{2,1+\lambda}}{\lambda}
 \longrightarrow m_2>0.
\]
We assume that $d\Lambda_\lambda$ has a  limit
$d\Lambda:=\lim_{\lambda \to 0}d\Lambda_\lambda$, with convergence of its fixed energy moments.
The moment expansion then gives
\begin{equation}
 \int_{\mathbb R}d\Lambda=m_2,\qquad
 \int_{\mathbb R}E^j\,d\Lambda(E)=m_{j+2},
 \qquad j\ge0.
 \label{eq:vn-limiting-inner-product-moments}
\end{equation}
The limiting orthogonality relation is therefore
\begin{equation}
 \int_{\mathbb R}q_j(E)q_l(E)\,d\Lambda(E)=\delta_{jl},
 \qquad q_0(E)=m_2^{-1/2}.
 \label{eq:vn-limiting-orthogonality}
\end{equation}
Consequently, the $q_j$ can be constructed by
orthogonalizing $1,E,E^2,\ldots$ with respect to
$d\Lambda$, with positive leading coefficients.
We assume that the resulting polynomials form a
complete basis in $L^2(d\Lambda)$.

With their normalization fixed, we now use these
polynomials to compute the leading amplitudes.
For $k\ge1$, orthogonality to the constant gives
\begin{equation}
 \varphi_{k,1+\lambda}(t)
 =\int_{\mathbb R}p_{k,1+\lambda}(E)
 \big(e^{-\ii Et}-1\big)\,d\nu_{1+\lambda}(E).
 \label{eq:vn-subtracted-amplitude}
\end{equation}
Substituting the leading polynomial form gives
\begin{align}
 \frac{\varphi_{k,1+\lambda}(t)}{\sqrt{\lambda}}
 ={}&
 \frac1\lambda\int_{\mathbb R}
 E q_{k-1}(E)\big(e^{-\ii Et}-1\big)
 \,d\nu_{1+\lambda}(E)
 \notag\\
 &+
 \frac{p_{k,1+\lambda}(0)}{\sqrt{\lambda}}
 \big[G_{1+\lambda}(t)-1\big]+o(1).
 \label{eq:vn-separated-scaled-amplitude}
\end{align}
The second line's explicit constant-term contribution
is $O(\lambda)$, since
$p_{k,1+\lambda}(0)=O(\sqrt{\lambda})$ and
$G_{1+\lambda}(t)-1=O(\lambda)$.
This is why the subtraction in
Eq.~\eqref{eq:vn-subtracted-amplitude} is useful:
it removes the constant term from the leading
rescaled amplitude.

We can now express the remaining integral using the
same measure that normalizes $q_j$:
\begin{equation}
 \frac{\varphi_{k,1+\lambda}(t)}{\sqrt{\lambda}}
 =
 \int_{\mathbb R}q_{k-1}(E)
 \frac{e^{-\ii Et}-1}{E}\,d\Lambda_\lambda(E)
 +o(1).
 \label{eq:vn-scaled-amplitude-measure}
\end{equation}
The quotient is regular at $E=0$, where its value is
$-\ii t$, and its absolute value is bounded by $|t|$.
Taking the replica limit at fixed $k$ and $t$ therefore
gives
\begin{equation}
 V_k(t)
 :=\lim_{\lambda\to 0}
 \frac{\varphi_{k,1+\lambda}(t)}{\sqrt{\lambda}}
 =
 \int_{\mathbb R}q_{k-1}(E)
 \frac{e^{-\ii Et}-1}{E}\,d\Lambda(E).
 \label{eq:vn-appendix-leading-amplitude}
\end{equation}
The continuous value at zero includes any contribution
from an atom of $d\Lambda$ there.
This determines the leading amplitude for each existing
limiting polynomial. The weighted sum required for the
complexity is considered below.

\subsubsection{The limiting polynomials and their recurrence}
The recurrence for $q_j$ follows from the original
Lanczos construction. In the spectral representation,
multiplication by $E$ represents the Hamiltonian,
and the finite-replica Krylov polynomials satisfy Eq.~\eqref{eq:spectral-three-term}.
We now identify the limit of this relation using
$\sqrt{\lambda}\,p_{j+1,1+\lambda}\to E q_j$.

The diagonal and off-diagonal Lanczos coefficients are
matrix elements of multiplication by $E$.
Using the leading polynomial forms and the measure
$d\Lambda$, their limits are
\begin{align}
 \alpha_j
 &:=\lim_{\lambda\to0}a_{j+1,1+\lambda}
 =\int_{\mathbb R}E q_j(E)^2\,d\Lambda(E),
 \notag\\
 \beta_{j+1}
 &:=\lim_{\lambda\to0}b_{j+2,1+\lambda}
 =\int_{\mathbb R}E q_{j+1}(E)q_j(E)\,d\Lambda(E).
 \label{eq:vn-lanczos-coefficient-limits}
\end{align}
These relations apply at fixed degrees before the limiting
polynomial sequence terminates.

Setting $k=j+1$ in Eq.~\eqref{eq:spectral-three-term},
multiplying by $\sqrt{\lambda}$, and taking the limit gives
\[
 E^2q_j(E)
 =E\big[
 \beta_{j+1}q_{j+1}(E)
 +\alpha_jq_j(E)
 +\beta_jq_{j-1}(E)
 \big].
\]
At $j=0$, the original term involving $p_0=1$ vanishes,
since $\sqrt{\lambda}\,b_{1,1+\lambda}=O(\lambda)$.
Cancelling the common polynomial factor $E$ therefore yields
\begin{equation}
 E q_j(E)
 =\beta_{j+1}q_{j+1}(E)
  +\alpha_jq_j(E)+\beta_jq_{j-1}(E),
 \qquad q_{-1}=0,\quad \beta_0=0.
 \label{eq:vn-limiting-lanczos}
\end{equation}
Thus this recurrence describes the limiting coefficients
of the original Krylov chain starting at site $k=1$.
In particular, $\alpha_0$ comes from $a_{1,n}$ and
$\beta_1$ comes from $b_{2,n}$, while the original first
link $b_{1,n}$ vanishes in the replica limit.

The same three-term structure can be seen directly from
the orthogonality of $q_j$.
For $l\le j-2$,
\[
 \int_{\mathbb R}q_l(E)E q_j(E)\,d\Lambda(E)
 =
 \int_{\mathbb R}[E q_l(E)]q_j(E)\,d\Lambda(E)=0,
\]
because $E q_l$ has degree below $j$.
Hence $E q_j$ has projections only onto
$q_{j-1}$, $q_j$, and $q_{j+1}$.
Symmetry of the inner product makes the coefficient of
$q_{j-1}$ equal to $\beta_j$, as determined at the
preceding step.

For practical computation, start with
\[
 q_{-1}(E)=0,\qquad
 q_0(E)=m_2^{-1/2},\qquad
 \beta_0=0.
\]
Subtracting the two known projections and normalizing
gives the recursive algorithm
\begin{align}
 \alpha_j
 &=\int_{\mathbb R}E q_j(E)^2\,d\Lambda(E),
 \notag\\
 \beta_{j+1}^{\,2}
 &=\int_{\mathbb R}
 \big[(E-\alpha_j)q_j(E)-\beta_jq_{j-1}(E)\big]^2
 \,d\Lambda(E),
 \notag\\
 q_{j+1}(E)
 &=\frac{(E-\alpha_j)q_j(E)-\beta_jq_{j-1}(E)}
         {\beta_{j+1}}.
 \label{eq:vn-polynomial-recursion}
\end{align}
Choosing $\beta_{j+1}>0$ preserves the positive
leading-coefficient convention.

All integrals in this recursion can be evaluated from
$\int E^a\,d\Lambda=m_{a+2}$.
The first step gives
\begin{equation}
 \alpha_0=\frac{m_3}{m_2},\qquad
 \beta_1^2=\frac{m_4}{m_2}
             -\left(\frac{m_3}{m_2}\right)^2,\qquad
 q_1(E)=\frac{E-\alpha_0}{\sqrt{m_2}\,\beta_1}.
 \label{eq:vn-polynomial-first-step}
\end{equation}
Higher polynomials follow by repeating the same procedure.
If $\beta_{j+1}=0$, the construction terminates and no
division by this coefficient is performed.

The resulting polynomials enter the leading amplitudes
through Eq.~\eqref{eq:vn-appendix-leading-amplitude}.
It remains to obtain their measure $d\Lambda$ and its
moments from the first-order return function $\Psi(t)$.

\subsubsection{Obtaining the measure from $\Psi(t)$}
To complete the calculation, we determine $d\Lambda$
from the first-order return function $\Psi(t)$.
The factor $E^2$ in this measure is generated by taking
two time derivatives of the spectral representation:
\begin{equation}
 -\frac{1}{\lambda}\,\partial_t^2G_{1+\lambda}(t)
 =
 \int_{\mathbb R}e^{-\ii Et}
 \frac{E^2}{\lambda}\,d\nu_{1+\lambda}(E)
 =
 \int_{\mathbb R}e^{-\ii Et}\,d\Lambda_\lambda(E).
 \label{eq:vn-prelimit-curvature-transform}
\end{equation}
Using $G_{1+\lambda}(t)=1-\lambda\Psi(t)+o(\lambda)$
under the spectral convergence assumptions stated above,
the replica limit gives
\begin{equation}
 \Psi''(t)
 =\int_{\mathbb R}e^{-\ii Et}\,d\Lambda(E).
 \label{eq:vn-appendix-curvature-transform}
\end{equation}
Thus $\Psi''$ is the Fourier transform of $d\Lambda$.
Knowing its full time dependence determines the measure,
including any discrete spectral contributions.

For the polynomial recursion, we can instead obtain
the required moments directly by differentiating at
$t=0$:
\begin{equation}
 \int_{\mathbb R}E^j\,d\Lambda(E)
 =\left.\ii^j\partial_t^{j+2}\Psi(t)\right|_{t=0}
 =m_{j+2},\qquad j\ge0.
 \label{eq:vn-appendix-curvature-moments}
\end{equation}
These moments determine the coefficients $\alpha_j$,
$\beta_j$, and the polynomials $q_j$ through
Eq.~\eqref{eq:vn-polynomial-recursion}.
They therefore provide the input needed to evaluate
the leading amplitudes.

For completeness, integrating
Eq.~\eqref{eq:vn-appendix-curvature-transform} twice,
with $\Psi(0)=0$ and $\Psi'(0)=\ii m_1$, gives
\begin{equation}
 \Psi(t)=\ii m_1t+
 \int_{\mathbb R}
 \frac{1-e^{-\ii Et}-\ii Et}{E^2}\,d\Lambda(E).
 \label{eq:vn-return-from-curvature}
\end{equation}
The quotient takes its continuous value $t^2/2$
at $E=0$.
The coefficient $m_1$ contributes only the linear
phase term, while $d\Lambda$ determines the leading
Krylov amplitudes.

\subsubsection{Summary for the Krylov-von Neumann complexity computing}
For $m_2=\Psi''(0)>0$, we briefly summarize the steps
for computing the Krylov-von Neumann complexity from the
first-order return function $\Psi(t)$.

First, determine the positive measure $d\Lambda$
from its Fourier transform,
\[
 \Psi''(t)=\int_{\mathbb R}e^{-\ii Et}\,d\Lambda(E).
\]
The moments required for the polynomial construction
can also be obtained directly:
\[
 \int_{\mathbb R}E^j\,d\Lambda(E)
 =\ii^j\Psi^{(j+2)}(0)=m_{j+2}.
\]
Next, starting with $q_0=m_2^{-1/2}$ and $\beta_0=0$,
use Eq.~\eqref{eq:vn-polynomial-recursion} to compute
the limiting Lanczos coefficients and the polynomials
$q_j$. Their overlaps with the time-dependent kernel
give the leading amplitudes,
\[
 V_{j+1}(t)
 =
 \int_{\mathbb R}q_j(E)
 \frac{e^{-\ii Et}-1}{E}\,d\Lambda(E),
\]
where the quotient takes its continuous value
$-\ii t$ at $E=0$.

Finally, substituting these amplitudes into the
definition of the complexity gives
\begin{align}
 \mathcal C_{\mathrm{vN}}(t)
 &=\sum_{j\ge0}(j+1)|V_{j+1}(t)|^2
 \notag\\
 &=\sum_{j\ge0}(j+1)
 \left|
 \int_{\mathbb R}q_j(E)
 \frac{e^{-\ii Et}-1}{E}\,d\Lambda(E)
 \right|^2.
 \label{eq:vn-appendix-complexity-result}
\end{align}
Here we assume that the weighted sum converges and
that it can be interchanged with the replica limit.
If the limiting polynomial sequence terminates,
the sum runs only over the existing polynomials.

A consistency check follows by summing the leading
probabilities without the Krylov weight.
Completeness of $q_j$ gives
\begin{align}
 \sum_{j\ge0}|V_{j+1}(t)|^2
 &=
 \int_{\mathbb R}
 \left|\frac{e^{-\ii Et}-1}{E}\right|^2d\Lambda(E)
 \notag\\
 &=2\int_{\mathbb R}
 \frac{1-\cos(Et)}{E^2}\,d\Lambda(E)
 =2\operatorname{Re}\Psi(t),
 \label{eq:vn-escape-weight}
\end{align}
where the last equality follows from
Eq.~\eqref{eq:vn-return-from-curvature}.
This agrees with the total probability of leaving
the initial Krylov site:
\[
 \sum_{k\ge1}|\varphi_{k,1+\lambda}(t)|^2
 =1-|G_{1+\lambda}(t)|^2
 =2\lambda\operatorname{Re}\Psi(t)+o(\lambda).
\]
Thus the limiting amplitudes account for the full
leading escape probability, including when the
limiting polynomial sequence terminates.

If $m_2=0$, positivity implies $d\Lambda=0$.
Equation~\eqref{eq:vn-return-from-curvature} then reduces
to $\Psi(t)=\ii m_1t$, so the leading escape probability
vanishes. Under the same assumption allowing the replica
limit to pass through the weighted sum, this gives
$\mathcal C_{\mathrm{vN}}(t)=0$.
No normalized polynomial construction is needed in
this case.

\subsection{Short-time expansion}
\label{app:short-time-expansion}

We derive the expansions in
Sec.~\ref{sec:short-time-expansion} directly from the spectral
representation. The calculation separates the probability
of leaving the initial state from the additional contribution
associated with reaching higher Krylov levels.

For finite chains, the Taylor expansions below are automatic.
For infinite chains, we assume sufficient spectral regularity
to expand the weighted probability sum.
A sufficient condition is a finite absolute exponential
moment of $d\nu_n$ for the finite-replica calculation, and
of $d\Lambda$ for the von Neumann calculation.
These conditions ensure that the weighted contribution
from levels $k\ge3$ is $O(t^6)$.

%\paragraph{Finite replica number.}

Let's start with
\[
 \varphi_{k,n}(t)
 =\int_{\mathbb R}p_{k,n}(E)e^{-\ii Et}\,d\nu_n(E).
\]
Because $p_{k,n}$ is orthogonal to every polynomial of
degree less than $k$, the first nonvanishing term in
the exponential expansion is of order $t^k$:
\begin{equation}
 \varphi_{k,n}(t)
 =\frac{(-\ii t)^k}{k!}
   \int_{\mathbb R}p_{k,n}(E)E^k\,d\nu_n(E)
   +O(t^{k+1}).
 \label{eq:short-time-amplitude-order}
\end{equation}
The polynomials and measure are real, so
$\varphi_{k,n}(-t)=\varphi_{k,n}(t)^*$.
Each site probability is therefore even in $t$;
this property does not require time-reversal symmetry
of the underlying system.

Using $\varphi_{0,n}=G_n$ and probability conservation,
we can write the complexity exactly as
\begin{equation}
 \mathcal C_n(t)
 =1-|G_n(t)|^2+
 \sum_{k\ge2}(k-1)|\varphi_{k,n}(t)|^2.
 \label{eq:short-time-escape}
\end{equation}
Equation~\eqref{eq:short-time-amplitude-order} then gives
Eq.~\eqref{eq:short-time-second-site}.
To evaluate it through order $t^4$, we need the return
probability through that order and only the leading
$t^2$ term in $\varphi_{2,n}$.

At fixed $n$, temporarily omit the replica index on the
cumulants. Their defining expansion is
\begin{equation}
 \log G_n(t)
 =-\ii\kappa_1t-\frac{\kappa_2}{2}t^2
  +\frac{\ii\kappa_3}{6}t^3
  +\frac{\kappa_4}{24}t^4+O(t^5).
 \label{eq:short-time-log-return}
\end{equation}
Taking twice the real part gives
\[
 \log|G_n(t)|^2
 =-\kappa_2t^2+\frac{\kappa_4}{12}t^4+O(t^6).
\]
Exponentiating therefore yields
\begin{equation}
 1-|G_n(t)|^2
 =\kappa_2t^2-
 \left(\frac{\kappa_2^2}{2}
       +\frac{\kappa_4}{12}\right)t^4
 +O(t^6).
 \label{eq:short-time-return-probability}
\end{equation}

We next compute the leading second-site amplitude.
For $\kappa_2>0$, the first nonconstant polynomial is
\[
 p_{1,n}(E)=\frac{E-\kappa_1}{\sqrt{\kappa_2}}.
\]
The central moments satisfy
\begin{align}
 \int_{\mathbb R}(E-\kappa_1)^2\,d\nu_n(E)
 &=\kappa_2,
 \notag\\
 \int_{\mathbb R}(E-\kappa_1)^3\,d\nu_n(E)
 &=\kappa_3,
 \notag\\
 \int_{\mathbb R}(E-\kappa_1)^4\,d\nu_n(E)
 &=\kappa_4+3\kappa_2^2.
 \label{eq:short-time-central-moments}
\end{align}
Thus the projection of $(E-\kappa_1)^2$ onto the constant
is $\kappa_2$, while its projection onto $p_{1,n}$ is
$\kappa_3/\sqrt{\kappa_2}$.
Subtracting both projections and normalizing gives
\begin{equation}
 p_{2,n}(E)
 =
 \frac{
 (E-\kappa_1)^2-\kappa_2
 -\dfrac{\kappa_3}{\kappa_2}(E-\kappa_1)
 }{
 \sqrt{\kappa_4+2\kappa_2^2-\kappa_3^2/\kappa_2}
 }.
 \label{eq:short-time-second-polynomial}
\end{equation}
The squared normalization follows from
Eq.~\eqref{eq:short-time-central-moments}.
If it vanishes, the chain terminates at $k=1$ and the
second-site contribution is absent.

The numerator of Eq.~\eqref{eq:short-time-second-polynomial}
is monic, so its difference from $E^2$ has degree at most one.
Orthogonality and normalization consequently give
\[
 \int_{\mathbb R}p_{2,n}(E)E^2\,d\nu_n(E)
 =\sqrt{\kappa_4+2\kappa_2^2-\frac{\kappa_3^2}{\kappa_2}}.
\]
Using Eq.~\eqref{eq:short-time-amplitude-order}, we obtain
\begin{equation}
 |\varphi_{2,n}(t)|^2
 =\frac14
 \left(
 \kappa_4+2\kappa_2^2-\frac{\kappa_3^2}{\kappa_2}
 \right)t^4+O(t^6).
 \label{eq:short-time-second-probability}
\end{equation}

Adding Eqs.~\eqref{eq:short-time-return-probability}
and \eqref{eq:short-time-second-probability},
the quartic coefficient simplifies to
\[
 -\frac{\kappa_2^2}{2}-\frac{\kappa_4}{12}
 +\frac14\left(
 \kappa_4+2\kappa_2^2-\frac{\kappa_3^2}{\kappa_2}
 \right)
 =\frac{\kappa_4}{6}
  -\frac{\kappa_3^2}{4\kappa_2}.
\]
This proves Eq.~\eqref{eq:short-time-cumulants}.
The result remains valid when the second site is absent,
provided $\kappa_2>0$.

Finally, the entropy definition
in Eq.~\eqref{eq:short-time-entropy-data} gives
\[
 \log G_n(t)^*=-(n-1)\big[R_n(t)+\ii I_n(t)\big].
\]
Comparing with the complex conjugate of
Eq.~\eqref{eq:short-time-log-return} fixes the signs:
\begin{align}
 \kappa_{1,n}&=-(n-1)I_n'(0),
 &
 \kappa_{2,n}&=(n-1)R_n''(0),
 \notag\\
 \kappa_{3,n}&=(n-1)I_n'''(0),
 &
 \kappa_{4,n}&=-(n-1)R_n^{(4)}(0).
 \label{eq:short-time-entropy-cumulants}
\end{align}
Substitution gives Eq.~\eqref{eq:short-time-renyi}.

%\paragraph{The von Neumann limit.}

We now apply the same calculation to the limiting amplitudes
derived in Appendix~\ref{app:vn-spectral-limit},
\[
 V_k(t)
 =\int_{\mathbb R}q_{k-1}(E)
   \frac{e^{-\ii Et}-1}{E}\,d\Lambda(E),
 \qquad k\ge1.
\]
We retain the replica-limit assumptions stated there.
For $m_2>0$, the polynomials obey
$q_0=m_2^{-1/2}$ and
$\int q_jq_l\,d\Lambda=\delta_{jl}$.

The expansion
\[
 \frac{e^{-\ii Et}-1}{E}
 =-\ii t-\frac{Et^2}{2}
  +\frac{\ii E^2t^3}{6}+\cdots
\]
shows that $V_k(t)=O(t^k)$:
all preceding terms vanish by orthogonality of $q_{k-1}$.
The probabilities $|V_k(t)|^2$ are again even in $t$.
Therefore,
\begin{equation}
 \mathcal C_{\mathrm{vN}}(t)
 =\sum_{k\ge1}|V_k(t)|^2+|V_2(t)|^2+O(t^6).
 \label{eq:short-time-vn-decomposition}
\end{equation}

Completeness gives the total leading escape weight,
\begin{align}
 \sum_{k\ge1}|V_k(t)|^2
 &=\int_{\mathbb R}
   \frac{2(1-\cos Et)}{E^2}\,d\Lambda(E)
 \notag\\
 &=m_2t^2-\frac{m_4}{12}t^4+O(t^6),
 \label{eq:short-time-vn-escape}
\end{align}
where $m_r=\int E^{r-2}\,d\Lambda$ for $r\ge2$.
The integrand takes its continuous value $t^2$ at $E=0$.

To compute $V_2$, orthogonalize $E$ against $q_0$.
Its constant projection is $m_3/m_2$, and the squared norm
after subtraction is
\[
 \int_{\mathbb R}
 \left(E-\frac{m_3}{m_2}\right)^2d\Lambda(E)
 =m_4-\frac{m_3^2}{m_2}.
\]
Hence, when this norm is positive,
\begin{equation}
 q_1(E)
 =\frac{E-m_3/m_2}{\sqrt{m_4-m_3^2/m_2}},
 \qquad
 \int_{\mathbb R}E q_1(E)\,d\Lambda(E)
 =\sqrt{m_4-\frac{m_3^2}{m_2}}.
 \label{eq:short-time-vn-first-polynomial}
\end{equation}
The constant term in the kernel does not contribute to
$V_2$, so
\begin{align}
 V_2(t)
 &=-\frac{t^2}{2}
   \int_{\mathbb R}E q_1(E)\,d\Lambda(E)+O(t^3),
 \notag\\
 |V_2(t)|^2
 &=\frac14\left(m_4-\frac{m_3^2}{m_2}\right)t^4
   +O(t^6).
 \label{eq:short-time-vn-second-probability}
\end{align}
If the norm vanishes, the limiting sequence contains only
$q_0$ and this contribution is zero.

Substituting Eqs.~\eqref{eq:short-time-vn-escape}
and \eqref{eq:short-time-vn-second-probability} into
Eq.~\eqref{eq:short-time-vn-decomposition} gives the
quartic coefficient
\[
 -\frac{m_4}{12}
 +\frac14\left(m_4-\frac{m_3^2}{m_2}\right)
 =\frac{m_4}{6}-\frac{m_3^2}{4m_2},
\]
and establishes Eq.~\eqref{eq:short-time-vn} directly
from the limiting spectral construction.

When $m_2=0$, positivity implies $d\Lambda=0$, so all
leading escape amplitudes vanish.
Under the same assumptions permitting the replica limit
to pass through the weighted sum,
$\mathcal C_{\mathrm{vN}}(t)=0$, as discussed in
Appendix~\ref{app:vn-spectral-limit}.

\section{Calculation details for the finite Dirac interval}
\label{app:dirac-analytic}

We supply the spectral calculation underlying
Sec.~\ref{sec:dirac-finite-interval}, using the same regulator
$\delta$, interval length $\ell$, and polynomial conventions.
We first reconstruct the finite-replica measure, then evaluate
the first-order amplitudes needed for $\mathcal C_{\mathrm{vN}}$.
The final two subsections discuss convergence and the late-time
growth.

\subsection{From the return amplitude to the finite-replica calculation}
\label{app:dirac-spectrum}

The starting point is
$\log G_n(t)=\gamma_n[f_\ell(\delta+\ii t)-f_\ell(\delta)]$.
To identify its spectral measure, first write
\begin{equation}
 f_\ell(s)=2\int_0^\infty e^{-sE}
 \frac{1-\cos(\ell E)}{E}\,dE,
 \qquad \operatorname{Re}s>0.
 \label{eq:dirac-log-integral}
\end{equation}
%This identity follows without an inverse-transform calculation:
%the derivative of the right-hand side with respect to $\ell$ is
%$2\int_0^\infty e^{-sE}\sin(\ell E)dE
%=2\ell/(s^2+\ell^2)$, equal to $\partial_\ell f_\ell(s)$,
%and both expressions vanish at $\ell=0$.
Subtracting the values at $s=\delta+\ii t$ and $s=\delta$
gives Eq.~\eqref{eq:dirac-jump}.

The density $j(E)$ represents the spectrum in $\log G_n$;
it is not yet the full energy density of the seed.
To obtain that density, exponentiate:
\begin{equation}
 G_n(t)=e^{-\gamma_nJ}\sum_{p=0}^\infty
 \frac{\gamma_n^p}{p!}
 \left[\int_0^\infty e^{-\ii Et}j(E)\,dE\right]^p.
 \label{eq:dirac-convolution-expansion}
\end{equation}
For $p\ge1$, the product of integrals integrates independent
variables $E_1,\ldots,E_p$, while its phase depends only on
their total energy $E=E_1+\cdots+E_p$. Collecting terms with
the same total energy gives
\begin{equation}
 \left[\int_0^\infty e^{-\ii Et}j(E)\,dE\right]^p
 =\int_0^\infty e^{-\ii Et}j^{*p}(E)\,dE,
 \qquad
 j^{*(p+1)}(E)=\int_0^E j^{*p}(x)j(E-x)\,dx.
 \label{eq:dirac-convolution-transform}
\end{equation}
The $p=0$ term is the constant $1$, whose spectral measure is
$\delta(E)dE$. Comparing with~\eqref{eq:spectral-measure} we can obtain the measure~\eqref{eq:dirac-measure}.
%Each convolution is nonnegative and has integral $J^p$, so
%the total spectral weight is
%$e^{-\gamma_nJ}\sum_{p\ge0}(\gamma_nJ)^p/p!=1$.
%The convolution index is a spectral expansion index; no
%identification with the number of physical fermions is needed.

For the Krylov calculation, it is convenient to extract the
energy cumulants before constructing the moments.
Replacing $-\ii t$ by $z$ in the preceding transform gives
\begin{equation}
 \int_0^\infty e^{zE}\,d\nu_n(E)
 =\exp\!\left[\gamma_n\int_0^\infty
 (e^{zE}-1)j(E)\,dE\right],
 \qquad \operatorname{Re}z<\delta.
 \label{eq:dirac-moment-generator}
\end{equation}
Differentiating its logarithm at $z=0$ yields
$\kappa_{m,n}=\gamma_n\int E^m j(E)dE$. with some calculations we have
%The integral is elementary: the exponential term gives
%$\Gamma(m)\delta^{-m}$ and the cosine gives the average
%of the same expression with $\delta\to\delta\pm\ii\ell$.
%Thus
\bea
 \kappa_{m,n}=2\gamma_n\Gamma(m)D_m(\delta),
 \quad
 D_m(s):=s^{-m}-\frac12
 \left[(s+\ii\ell)^{-m}+(s-\ii\ell)^{-m}\right],
 \quad m\ge1.
 \label{eq:dirac-cumulants}
\eea

To use these cumulants in the polynomial recursion, we need
the moments $\mu_{m,n}=\int E^m d\nu_n$.
Differentiating the exponential relation in
Eq.~\eqref{eq:dirac-moment-generator} once and matching powers
of $z$ gives the recursive conversion
\begin{equation}
 \mu_{0,n}=1,\qquad
 \mu_{m+1,n}
 =\sum_{a=0}^m\binom ma\kappa_{a+1,n}\mu_{m-a,n}.
 \label{eq:dirac-moment-recursion}
\end{equation}
For example, $\mu_{1,n}=\kappa_{1,n}$,
$\mu_{2,n}=\kappa_{2,n}+\kappa_{1,n}^2$, and
$\mu_{3,n}=\kappa_{3,n}+3\kappa_{1,n}\kappa_{2,n}
+\kappa_{1,n}^3$.
These moments evaluate every polynomial inner product in
Eq.~\eqref{eq:spectral-recursion}, thereby determining both
$p_{k,n}$ and the original Lanczos coefficients.
For example,
\begin{equation}
 p_{1,n}(E)=\frac{E-\kappa_{1,n}}{\sqrt{\kappa_{2,n}}},
 \qquad
 \varphi_{1,n}(t)
 =\frac{\ii\partial_tG_n(t)-\kappa_{1,n}G_n(t)}
 {\sqrt{\kappa_{2,n}}}.
 \label{eq:dirac-first-amplitude}
\end{equation}
Higher amplitudes follow from the same procedure and
Eq.~\eqref{eq:dirac-renyi-series}.
%Thus the convolution spectrum, cumulants, and polynomial
%recursion are successive steps in one calculation.

\subsection{Evaluating the von Neumann amplitudes}
\label{app:dirac-vn-calculation}

Only the first replica variation is required for
$\mathcal C_{\mathrm{vN}}$. Expanding the full measure at
$n=1+\lambda$ gives
\begin{equation}
 d\nu_{1+\lambda}(E)
 =\delta(E)dE+\frac c3\lambda
 \left[j(E)\,dE-J\delta(E)dE\right]+O(\lambda^2).
 \label{eq:dirac-first-order-measure}
\end{equation}
Multiplication by $E^2/\lambda$ removes the zero-energy term
and retains $(c/3)E^2j(E)dE$, precisely the measure
$d\Lambda$ in Eq.~\eqref{eq:dirac-curvature}.
This also verifies directly the result obtained from $\Psi''$.
%The positive regulator ensures that this expansion remains
%valid when integrated against any fixed energy polynomial.

The moments needed to construct $q_j$ are therefore
\begin{equation}
 m_a=\int_0^\infty E^{a-2}\,d\Lambda(E)
 =\frac{2c}{3}\Gamma(a)D_a(\delta),
 \qquad a\ge2.
 \label{eq:dirac-limiting-moments}
\end{equation}
Starting with $q_0=m_2^{-1/2}$, these moments determine
successive $q_j$ through Eq.~\eqref{eq:vn-polynomial-recursion}.
No new polynomial construction is needed for this example.

Once a polynomial has been obtained, its amplitude can be
evaluated without a numerical energy integral.
Write $q_j(E)=\sum_{a=0}^j q_{ja}E^a$ and substitute
Eq.~\eqref{eq:dirac-curvature} into
Eq.~\eqref{eq:dirac-vn-series}:
\begin{align}
 V_{j+1}(t)
 &=\frac{2c}{3}\sum_{a=0}^j q_{ja}
 \int_0^\infty E^a[1-\cos(\ell E)]
 \left[e^{-(\delta+\ii t)E}-e^{-\delta E}\right]dE
 \notag\\
 &=\frac{2c}{3}\sum_{a=0}^j q_{ja}\,a!
 \left[D_{a+1}(\delta+\ii t)-D_{a+1}(\delta)\right].
 \label{eq:dirac-vn-amplitudes}
\end{align}
%The second line uses the same elementary Laplace integrals
%as Eq.~\eqref{eq:dirac-cumulants}.
%For example,
%\begin{equation}
% V_1(t)=\frac{2c}{3\sqrt{m_2}}
% [D_1(\delta+\ii t)-D_1(\delta)].
% \label{eq:dirac-vn-first-amplitude}
%\end{equation}
Each amplitude thus has explicit time dependence; the full
complexity follows by summing $(j+1)|V_{j+1}|^2$.

%As a normalization check, completeness of the $q_j$ gives
%\begin{equation}
% \sum_{j\ge0}|V_{j+1}(t)|^2
% =\frac{2c}{3}\int_0^\infty[1-\cos(Et)]j(E)\,dE
 %=2\operatorname{Re}\Psi(t).
% \label{eq:dirac-escape-check}
%\end{equation}
%This agrees with the first-order probability of leaving
%the seed, obtained independently from $1-|G_{1+\lambda}|^2$.

The short-time result follows directly from
Eq.~\eqref{eq:short-time-vn} and Eq~\eqref{eq:dirac-limiting-moments}. The result is
\bea
 \mathcal C_{\mathrm{vN}}(t)
 =\frac{2c}{3}\left[d_2u^2+
 \left(d_4-\frac{d_3^2}{d_2}\right)u^4\right]+O(u^6),
 \label{eq:dirac-short-moments}
\eea
where $u=t/\delta$ and $d_a=\delta^aD_a(\delta)=1-\operatorname{Re}(1+\ii r)^{-a}$ with $r:=\ell/\delta$. This yields Eq.~\eqref{eq:dirac-short} in the main text.
%Here $d_2=r^2(r^2+3)/(1+r^2)^2$ and
%$d_4-d_3^2/d_2=r^2(r^2-6)/[(1+r^2)^2(r^2+3)]$,
%which yields Eq.~\eqref{eq:dirac-short}.
%Using Eq.~\eqref{eq:dirac-cumulants} in
%Eq.~\eqref{eq:short-time-cumulants} gives the finite-$n$
%version stated in the main text.

%The resulting series converges at every fixed real time.
%Indeed, the Euclidean regulator provides exponential
%suppression at high energy, while the continuous part of
%$d\nu_{1+\lambda}$ has weight $O(\lambda)$.
%Consequently, $\varphi_{k,1+\lambda}(z)$ is analytic and
%uniformly $O(\sqrt{\lambda})$ in
%$\operatorname{Im}z<b$, for $0<b<\delta/2$.
%Orthogonality also makes its Taylor expansion start at
%order $z^k$. Applying Schwarz's lemma in this half-plane gives
%\begin{equation}
% |\varphi_{k,1+\lambda}(t)|^2
% \le A_b\lambda
% \left(\frac{t^2}{t^2+4b^2}\right)^k,
% \qquad k\ge1,
% \label{eq:dirac-replica-bound}
%\end{equation}
%where $A_b$ is independent of $k$ and sufficiently small
%$\lambda>0$.
%For fixed $t$, this bound remains summable after multiplication
%by $k$ and division by $\lambda$.
%It therefore justifies taking the replica limit through
%the Krylov sum and establishes that the series obtained
%above equals $\mathcal C_{\mathrm{vN}}(t)$.

\subsection{Late-time spreading at fixed interval length}
\label{app:dirac-long-time}

We give a derivation of the leading large-time behavior of $\mathcal{C}_{\text{vN}}(t)$ for Dirac fermion example at fixed $\ell,\delta>0$.
These results also show why the Krylov--von Neumann complexity continues to
grow when the timelike entropy approaches a constant, and how
the interval length enters its leading growth coefficient.
Throughout this subsection, the replica limit is taken first,
followed by $t\to+\infty$ at fixed $\ell,\delta>0$.
We use a large-degree wave-packet approximation to the
polynomial expansion, with its assumptions stated below.

%\paragraph{The part of the amplitude that spreads.}
Our starting point is Eq.~\eqref{eq:dirac-vn-series}. 
The time dependence enters through the increasingly rapid
oscillations of $e^{-\ii Et}$. To follow their contribution
along the Krylov chain, separate the amplitude as
\begin{equation}
 V_{j+1}^{\mathrm{osc}}(t)
 :=\int_0^\infty q_j(E)\frac{e^{-\ii Et}}{E}\,d\Lambda(E).
 \label{eq:dirac-late-oscillating-part}
\end{equation}
It is obvious that $V_{j+1}(t)=V_{j+1}^{\mathrm{osc}}(t)-V_{j+1}^{\mathrm{osc}}(0)$.
We first check that the fixed subtraction contributes only
below order $t^2$. Use
$1-\cos(\ell E)\le\ell^2E^2/2$ and compare polynomial
approximation errors with those for the Laguerre weight
$E^3e^{-\delta E}$. This gives the bound
\begin{equation}
 \sum_{j\ge0}(j+1)|V_{j+1}^{\mathrm{osc}}(t)|^2
 \le \frac{2c}{3}r^2(1+u^2).
 \label{eq:dirac-late-oscillating-bound}
\end{equation}
Substituting
$V_{j+1}(t)=V_{j+1}^{\mathrm{osc}}(t)
-V_{j+1}^{\mathrm{osc}}(0)$
into the complexity gives
\bea
 \mathcal C_{\mathrm{vN}}(t)
 ={}
 \sum_{j\ge0}(j+1)|V_{j+1}^{\mathrm{osc}}(t)|^2
+\sum_{j\ge0}(j+1)|V_{j+1}^{\mathrm{osc}}(0)|^2
 -2\operatorname{Re}\sum_{j\ge0}(j+1)
 V_{j+1}^{\mathrm{osc}}(t)
 V_{j+1}^{\mathrm{osc}}(0)^*.\nn \\
 ~
 \label{eq:dirac-late-complexity-split}
\eea
The second term is finite by
Eq.~\eqref{eq:dirac-late-oscillating-bound}
and independent of time, so it contributes only $O(1)$.
The first term is at most $O(u^2)$ by the same bound.
Cauchy--Schwarz therefore bounds the magnitude of the
third term by $O(u)$.
Thus neither the second nor the third term contributes
at order $u^2$, and
\begin{equation}
 \mathcal C_{\mathrm{vN}}(t)
 =\sum_{j\ge0}(j+1)|V_{j+1}^{\mathrm{osc}}(t)|^2
 +O(u).
\end{equation}
We now estimate the leading contribution to the above formula.
For this purpose, we need both the Krylov degrees that
contribute at large time and the magnitude of their
amplitudes. We first calculate these quantities for the
reference Laguerre weight, and then state the approximation
used for the Dirac measure.

Consider the reference weight $w_0(E)=E e^{-\delta E}$.
Its orthonormal polynomials with positive leading
coefficients are
\bea
 q_j^{(0)}(E)
 =\frac{(-1)^j\delta}{\sqrt{j+1}}
 L_j^{(1)}(\delta E).
\eea
For large $j$ and energies in a fixed positive interval,
the Laguerre asymptotic formula gives~\cite{DLMFLaguerre}
\begin{equation}
 q_j^{(0)}(E)
 \simeq
 \frac{(-1)^j\delta^{1/4}}
 {\sqrt{\pi}\,j^{1/4}E^{1/4}\sqrt{w_0(E)}}
 \cos\left(2\sqrt{j\delta E}-\frac{3\pi}{4}\right).
 \label{eq:dirac-reference-polynomial-asymptotic}
\end{equation}
Both the oscillatory phase and the prefactor will enter
the amplitude calculation.

To apply this formula away from the endpoint $E=0$,
consider the contribution from a fixed energy interval
$[E_a,E_b]$, with $0<E_a<E_b$:
\begin{equation}
 V_{j+1}^{(0)}(t)
 :=
 \int_{E_a}^{E_b}
 q_j^{(0)}(E)\frac{e^{-\ii Et}}{E}w_0(E)\,dE.
\end{equation}
Writing the cosine as two exponentials gives the phases
$\pm2\sqrt{j\delta E}-Et$.
For $t>0$, only the positive-sign branch $\Phi_j(E):=2\sqrt{j\delta E}-Et$ has a stationary
point at positive energy:
\begin{equation}
 \Phi_j'(E_j^*)=0\ \Rightarrow
 E_j^*=\frac{j\delta}{t^2}.
 \label{eq:dirac-late-depth-energy}
\end{equation}
We take $t,j\to\infty$ with $j/t^2$ fixed, so that
$E_j^*$ remains in the interior of the chosen energy
interval.

The stationary-phase method evaluates the integral by
expanding $\Phi_j$ to quadratic order around $E_j^*$.
Using $\Phi_j''(E_j^*)=-\frac{t}{2E_j^*}$,
the resulting oscillatory Gaussian integral supplies
the factor~\cite{DLMFStationaryPhase}
\[
 \sqrt{\frac{2\pi}{|\Phi_j''(E_j^*)|}}
 =
 \sqrt{\frac{4\pi E_j^*}{t}}.
\]
Combining this factor with the prefactor in
Eq.~\eqref{eq:dirac-reference-polynomial-asymptotic},
including the factor $1/2$ from the cosine, gives
\begin{align}
 |V_{j+1}^{(0)}(t)|^2
 &\simeq
 \frac{\sqrt{\delta}}{4\pi\sqrt j}
 \frac{w_0(E_j^*)}{(E_j^*)^{5/2}}
 \frac{4\pi E_j^*}{t}
 \notag\\
 &=\frac{\delta}{t^2}
   \frac{w_0(E_j^*)}{(E_j^*)^2}.
 \label{eq:dirac-reference-probability}
\end{align}
Thus stationary phase determines not only the relevant
energy $E_j^*$, but also the probability prefactor
$\delta/t^2$.

For the actual Dirac measure, write
\[
 d\Lambda(E)=w(E)\,dE,
 \qquad
 w(E)=\frac{2c}{3}E e^{-\delta E}
                 [1-\cos(\ell E)].
\]
%We assume that the leading probability distribution
%retains the form of
%Eq.~\eqref{eq:dirac-reference-probability}, with the full
%Dirac density.
Motivated by the reference Laguerre calculation, we adopt
the following leading approximation for the Dirac
probabilities:
\begin{equation}
 |V_{j+1}^{\mathrm{osc}}(t)|^2
 \simeq
 \frac{\delta}{t^2}
 \frac{w(E_j^*)}{(E_j^*)^2},
 \qquad E_j^*=\frac{j\delta}{t^2}.
 \label{eq:dirac-late-packet-weight}
\end{equation}
This approximation is taken in the joint large-$t$,
large-$j$ limit, with $E_j^*$ fixed and away from the
zeros of $w(E)$.
It assumes that the reference stationary-phase result
extends to the leading probability magnitude when
$w_0(E)$ is replaced by the full Dirac density $w(E)$.
This is an additional assumption about the Dirac
polynomials, beyond the location of the stationary point.

The conversion of the Krylov sum into an energy integral
is now explicit. Consecutive stationary energies are
separated by
\begin{equation}
 \Delta E=E_{j+1}^*-E_j^*=\frac{\delta}{t^2}.
 \label{eq:dirac-late-energy-spacing}
\end{equation}
Using $j=t^2E_j^*/\delta$, the leading weighted sum becomes
\begin{align}
 \mathcal C_{\mathrm{vN}}(t)
 &\simeq
 \sum_{j\ge1}j\,
 \frac{\delta}{t^2}
 \frac{w(E_j^*)}{(E_j^*)^2}
 \notag\\
 &=\frac{t^2}{\delta}
 \sum_{j\ge1}
 \frac{w(E_j^*)}{E_j^*}\,\Delta E
 \notag\\
 &\simeq
 \frac{t^2}{\delta}
 \int_0^\infty\frac{w(E)}{E}\,dE.
 \label{eq:dirac-late-riemann-sum}
\end{align}
The last step is a Riemann-sum approximation.
Replacing $j+1$ by $j$ changes only a finite term.
In extending the calculation from fixed energy intervals
to the full sum, we also assume that levels far beyond
$j\sim u^2$ contribute no additional term of order $u^2$.

Finally, the remaining integral is elementary:
\bea
 \frac1\delta\int_0^\infty\frac{w(E)}{E}\,dE=
 \frac{2c}{3\delta}
 \int_0^\infty e^{-\delta E}
               [1-\cos(\ell E)]\,dE
 %\notag\\
% &=
 %\frac{2c}{3\delta}
 %\left(\frac1\delta-\frac{\delta}{\delta^2+\ell^2}\right)
% \notag\\
% &=
 =\frac{2c}{3}
 \frac{\ell^2}{\delta^2(\delta^2+\ell^2)}.
 \label{eq:dirac-late-coefficient}
\eea
Under the stated probability and tail assumptions,
we therefore obtain
\bea
 \mathcal C_{\mathrm{vN}}(t)
 =
 \frac{2c}{3}\frac{r^2}{1+r^2}u^2+o(u^2).
\eea
%\paragraph{Relation to the entropy plateau.}
The large-time coefficient can be expressed in terms of
the entropy plateau,
\[
 \Psi_\infty
 =\frac c3\log\left(1+\frac{\ell^2}{\delta^2}\right).
\]
Differentiating at fixed $\ell$ and $c$ gives
\begin{equation}
 -\left.\partial_\delta\Psi_\infty\right|_{\ell,c}
 =\frac{2c}{3}
 \left(\frac1\delta-\frac{\delta}{\delta^2+\ell^2}\right)
 =\int_0^\infty\frac{d\Lambda(E)}{E}.
 \label{eq:dirac-plateau-derivative}
\end{equation}
Combining this identity with the preceding large-time
approximation, we obtain
\begin{equation}
 \lim_{t\to\infty}
 \frac{\mathcal C_{\mathrm{vN}}(t)}{t^2}
 =\frac1\delta\int_0^\infty\frac{d\Lambda(E)}{E}
 =-\frac1\delta
 \left.\partial_\delta\Psi_\infty\right|_{\ell,c}.
 \label{eq:dirac-plateau-growth}
\end{equation}

\newpage
\bibliographystyle{JHEP}
\bibliography{references}

\end{document}